\documentclass[12pt]{spieman}  
\usepackage{amsmath,amsfonts,amssymb}
\usepackage{graphicx}
\usepackage{setspace}
\usepackage{tocloft}
\usepackage{changes} 
\setdeletedmarkup{ \color{gray}{(Deleted: #1)}}

\newcommand{\arcsec}{^{\prime\prime}}

\newcommand{\cref}[1]{\textsuperscript{\ref{#1}}}

\newcommand{\farcs}{\mbox{\ensuremath{.\!\!^{\prime\prime}}}}

\title{Pushing the Limits of the Coronagraphic Occulters on {\it HST}/STIS}

\author[a]{John H. Debes}
\author[b,c]{Bin Ren}
\author[d]{Glenn Schneider}
\affil[a]{Space Telescope Science Institute, 3700 San Martin Dr., Baltimore, MD, USA 21218}
\affil[b]{Department of Physics and Astronomy, The Johns Hopkins University, 3400 N.~Charles St., Baltimore, MD, USA 21218}
\affil[c]{Department of Applied Mathematics and Statistics, The Johns Hopkins University,  3400 N.~Charles St., Baltimore, MD, USA 21218}
\affil[d]{Steward Observatory and the Department of Astronomy, The University of Arizona, 933 North Cherry Avenue, Tucson, AZ 85721 USA}

\cftpagenumbersoff{figure}
\cftpagenumbersoff{table} 
\begin{document} 
\maketitle

\begin{abstract}
The {\it Hubble Space Telescope} ({\it HST})/Space Telescope Imaging Spectrograph (STIS) contains the only currently operating coronagraph in space that is not trained on the Sun. In an era of extreme--adaptive-optics--fed coronagraphs, and with the possibility of future space-based coronagraphs, we re-evaluate the contrast performance of the STIS CCD camera.  The 50CORON aperture consists of a series of occulting wedges and bars, including the recently commissioned BAR5 occulter.  We discuss the latest procedures in obtaining high contrast imaging of circumstellar disks and faint point sources with STIS. For the first time, we develop a noise model for the coronagraph, including systematic noise due to speckles, which can be used to predict the performance of future coronagraphic observations. Further, we present results from a recent calibration program that demonstrates better than $10^{-6}$ point-source contrast at $0\farcs6$, ranging to $3\times10^{-5}$ point-source contrast at $0\farcs25$.  These results are obtained by a combination of sub-pixel grid dithers, multiple spacecraft orientations, and post-processing techniques.  Some of these same techniques will be employed by future space-based coronagraphic missions. We discuss the unique aspects of STIS coronagraphy relative to ground-based adaptive-optics--fed coronagraphs.

\end{abstract}

\keywords{image processing,speckle,point spread functions,planets,charge-coupled-device imagers}

{\noindent \footnotesize\textbf{*}John H. Debes,  \linkable{debes@stsci.edu} }

\begin{spacing}{2}   

\section{Introduction}
\label{sect:intro}  
STIS is a second generation instrument for {\it HST} that was installed in early 1997\cite{woodgate98} during Servicing Mission Two.  It possesses a comprehensive collection of imaging and spectroscopic modes that span from the far ultraviolet to the near infrared (IR).  It is currently the oldest operating instrument on board {\it HST} that primarily obtains science observations.  It retains a significant proportion of science time on {\it HST}, with 19\% of all General Observer (GO) orbits during Cycles 23--25.

\begin{figure}
\begin{center}
\includegraphics[width=0.6\textwidth]{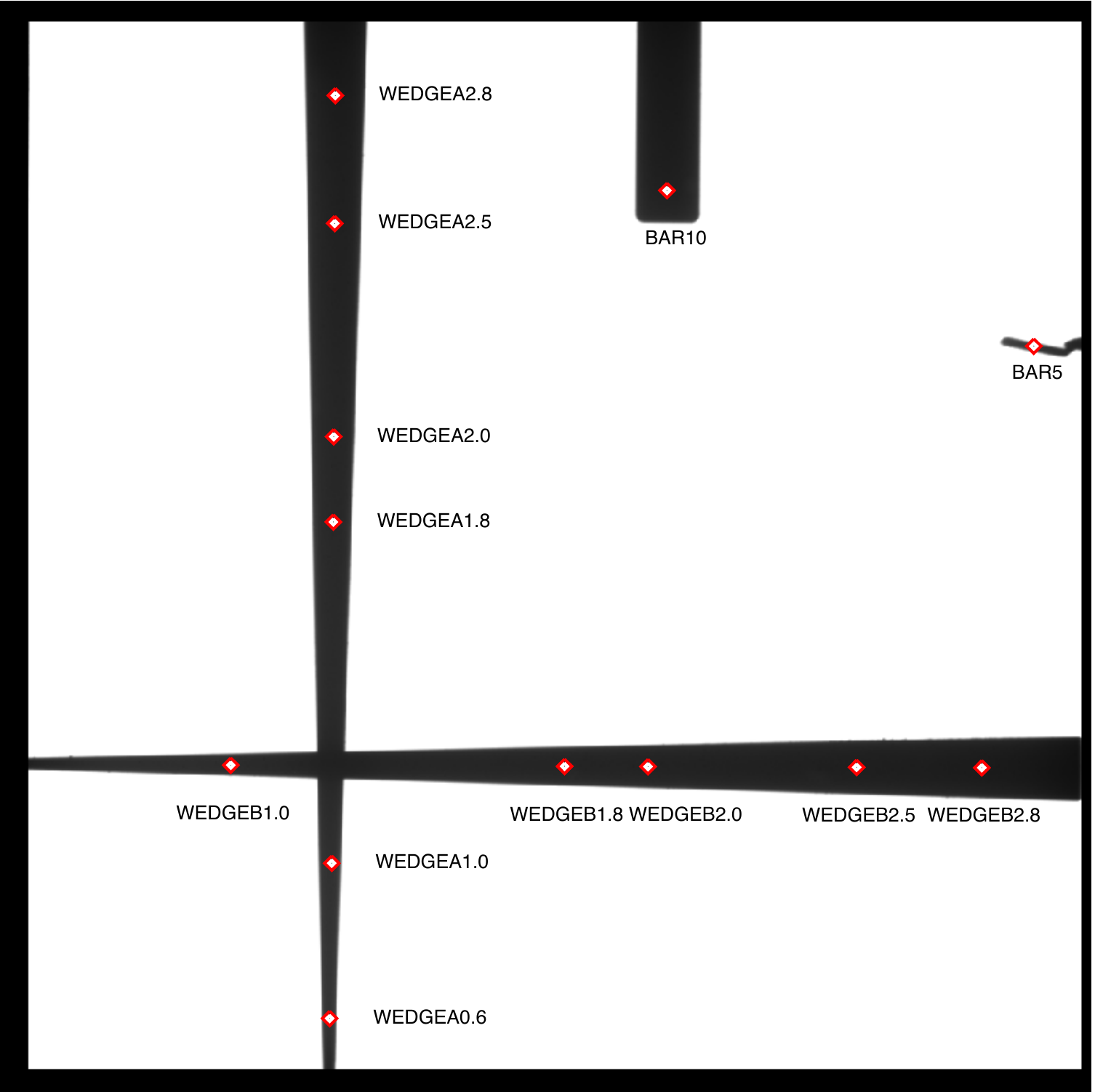}  
\end{center}
\caption 
[The 50CORON aperture mask with supported aperture locations.]{ \label{fig:f1} The 50CORON aperture mask, it is designed to have a full field of view of $52\arcsec\times52\arcsec$ over $1024\times1024$ pixels (${\sim}0\farcs05078$ pixel$^{-1}$; resolution element $\sim$0\farcs07). Red diamonds denote supported aperture locations within the 50CORON mask. The STIS coronagraphic mode is unfiltered, allowing light over the full wavelength range of sensitivity for the CCD, i.e., from ${\sim}0.20$ $\mu$m to ${\sim}1.03$ $\mu$m, with a pivot wavelength $\lambda_{\rm c}=0.5852$ $\mu$m and $\Delta\lambda/\lambda_{\rm c}\approx75\%$  (cf.~Chapter 5 of Ref.~\citenum{stisihb18}).
} 
\end{figure}

The high contrast imaging mode on STIS includes a focal plane mask designated 50CORON. The focal plane mask consists of two wedges (A and B) joined perpendicularly, along with a wide occulting bar at the top of the field-of-view (FOV), as projected on-sky in {\it HST}'s Science Aperture Instrument Frame\footnote{\label{note:siaf}\url{http://www.stsci.edu/hst/observatory/apertures/siaf.html
}}, and a narrow occulting finger on the right edge of the FOV (See Figure \ref{fig:f1}).  On one hand, positions on the wedges are designated according to the width of the wedge at that location: for example, WEDGEA1.0 corresponds to the location on the vertical wedge with a width of $1\farcs0$. On the other hand, the other apertures have designed lengths in the names: a wide occulting bar named BAR10 is $10\arcsec$ long, and an occulting finger denoted BAR5 has a length of $5\arcsec$.  BAR5 was bent during the assembly of STIS, and was only recently commissioned for high contrast imaging via the outsourced calibration Program 12923 (PI: A. G\'asp\'ar)\cite{schneider17}. It is now a fully supported aperture location for STIS. The occulters are combined with a hard-edged Lyot stop in the pupil plane that is present for all imaging modes of STIS. The combination of the occulters and Lyot stop creates a simple coronagraph that marginally suppresses the diffraction pattern of the telescope\cite{grady03}.

In this article, we investigate the limits to STIS' performance with respect to total-intensity, visible-light, high-contrast imaging. We determine what sets its limits on contrast performance at small inner working angles; in addition, we can assess its implications for what might be possible with second-generation high-contrast imagers in space, e.g., the coronagraphic instrument (CGI) proposed for the {\it Wide Field Infrared Survey Telescope} ({\it WFIRST}) mission\cite{trauger16,douglas18}. In Section \ref{sec:hi-c} we talk about general strategies to obtain high contrast with STIS, building upon previous results\cite{krist04,grady05,schneider14}; therein we also include the development of a coronagraphic noise model, which can be used to predict the performance of a given observing strategy; together with an analysis of the impact that charge transfer inefficiency has on high contrast imaging with CCDs. In Section \ref{sec:push} we present a calibration program recently obtained with BAR5 to demonstrate high-contrast capabilities with STIS that rival current ground-based near-infrared--optimized coronagraphs which are equipped with extreme adaptive optics (AO) systems, as well as a discussion of STIS' sensitivity to point sources and circumstellar disks. In Section \ref{sec:cases} we describe a handful of common science cases that may require STIS' coronagraphic aperture and the considerations required. In Section \ref{sec:wfirst}, we apply our new STIS coronagraphic noise model to investigate the potential of {\it WFIRST} to conduct high contrast imaging beyond its dark hole, and in Section \ref{sec:advantage} we list the areas in which STIS is unique compared to other high contrast imagers. We finish with our conclusions in Section \ref{sec:conc}. 

\section{High Contrast Imaging with STIS}
\label{sec:hi-c}
In their simplest form, high contrast imaging techniques seek to minimize the diffracted light from a brighter central astrophysical object, allowing fainter point sources or extended sources to be detected at small angular separations from the host object. 


High contrast imaging with STIS requires two major steps to accomplish this task. The first is to obtain the image of a target, which is achieved by placing the target behind the occulters---this prevents light from the core of the stellar point spread function (PSF) from saturating and bleeding on the CCD detector. The raw contrast in such images is typically not sufficient for most observations, since a significant halo of the PSF (also called the wings of the PSF) is present. The second step is to subtract off an astrometrically co-registered, and intensity scaled reference PSF to enhance contrast. A final optional step is to obtain additional images of the target at different spacecraft orientations to obtain information about the sky scene that is occulted by the wedges, bars, and diffraction spikes of the target.

High contrast imaging programs with STIS in the first few years of its operation primarily focused on a strategy of obtaining images of a target at two separate spacecraft orientations and with a single observation of a reference star\cite{grady03}.  This strategy minimizes the total number of {\it HST} orbits required to obtain a final sky image (3), but at the expense of angular coverage, inner working angle, and exposure depth at large angles.  In this observing scheme, the reference star for the target must also be very closely matched in spectral type (or spectral energy distribution, SED): the ideal target and PSF reference star (a.k.a.~PSF-template star) should have a difference in $|B-V|$ or $|V-I|$ colors smaller than $0.05$. These early programs found that the contrast achieved with a given aperture location is the same--thus paving the way for combining aperture positions to maximize angular coverage or sensitivity. The 50CORON aperture mode is operated without any filters and the effective bandpass of the aperture is defined by the red and blue wing sensitivity cut-offs of the CCD detector. The broad effective bandpass thus creates different PSF wings as a function of the host star's SED. Other high contrast techniques were also investigated, such as coronagraphic spectroscopy with the STIS long slit gratings \cite{roberge05}, and placing bright sources outside of the field of view to attempt deep imaging of nearby stars, such as $\epsilon$~Eri \cite{proffitt04}.

With the refurbishment of STIS in 2009 during the fourth {\it HST} servicing mission (SM4)\cite{rinehart08,goudfrooij09}, STIS became the sole space-based coronagraph in operation on {\it HST}, since ACS/HRC was not repaired during the mission.  Since SM4, multiple programs have  used the STIS 50CORON aperture to obtain high contrast at large separations to directly image and characterize the orbit of Fomalhaut~b, to directly image at high signal-to-noise debris disks previously unobserved in the visible, and to find very faint disks \cite{kalas13, schneider14,krist12}. Many of these programs adopted observations at multiple spacecraft orientations (known as multi-roll differential imaging, MRDI), and some used the central star itself as a reference, akin to ground based angular differential imaging (ADI) techniques \cite{marois08,macintosh15}.

Concurrently, post-processing techniques that suppress speckle noise, such as the Locally Optimized Combination of Images (LOCI) algorithm \cite{lafreniere07}, the Karheunen-Lo\`eve Image Projection (KLIP) algorithm \cite{soummer12}, and Non-negative Matrix Factorization \cite{ren18} have been developed and shown to be successful at pushing contrasts closer to the photon noise limit of a given coronagraph.  These techniques have been shown to be effective with archival NICMOS images\cite{choquet14, choquet17, hagan18, ren18b} from its calibrated archive\footnote{\url{https://archive.stsci.edu/prepds/laplace/index.html}}, and STIS\cite{ren17} coronagraphic images.

\subsection{Strategies that Maximize Angular Coverage}

There exist several strategies for obtaining full $360^\circ$ circum-azimuthal field coverage of an object with STIS, but given the multiple options, we provide a subset of possibilities to give observers a flavor of what is possible. We detail some common examples in Section \ref{sec:cases}. Wedge and BAR position selection is primarily dictated by the desired inner working angle and the brightness of the star.  However, due to the occulting mask obscurations and the presence of unapodized stellar diffraction spikes in the image scene, a significant portion of the stellocentric field is unimaged in a single spacecraft orientation. This can be mitigated by using multiple spacecraft orientations and/or by choosing a combination of aperture positions; see Ref.~\citenum{apai15} for the observation of the $\beta$~Pic disk with 3 roll angles and 2 aperture positions on each wedge.

\begin{figure}
\begin{center}
\includegraphics[width=\textwidth]{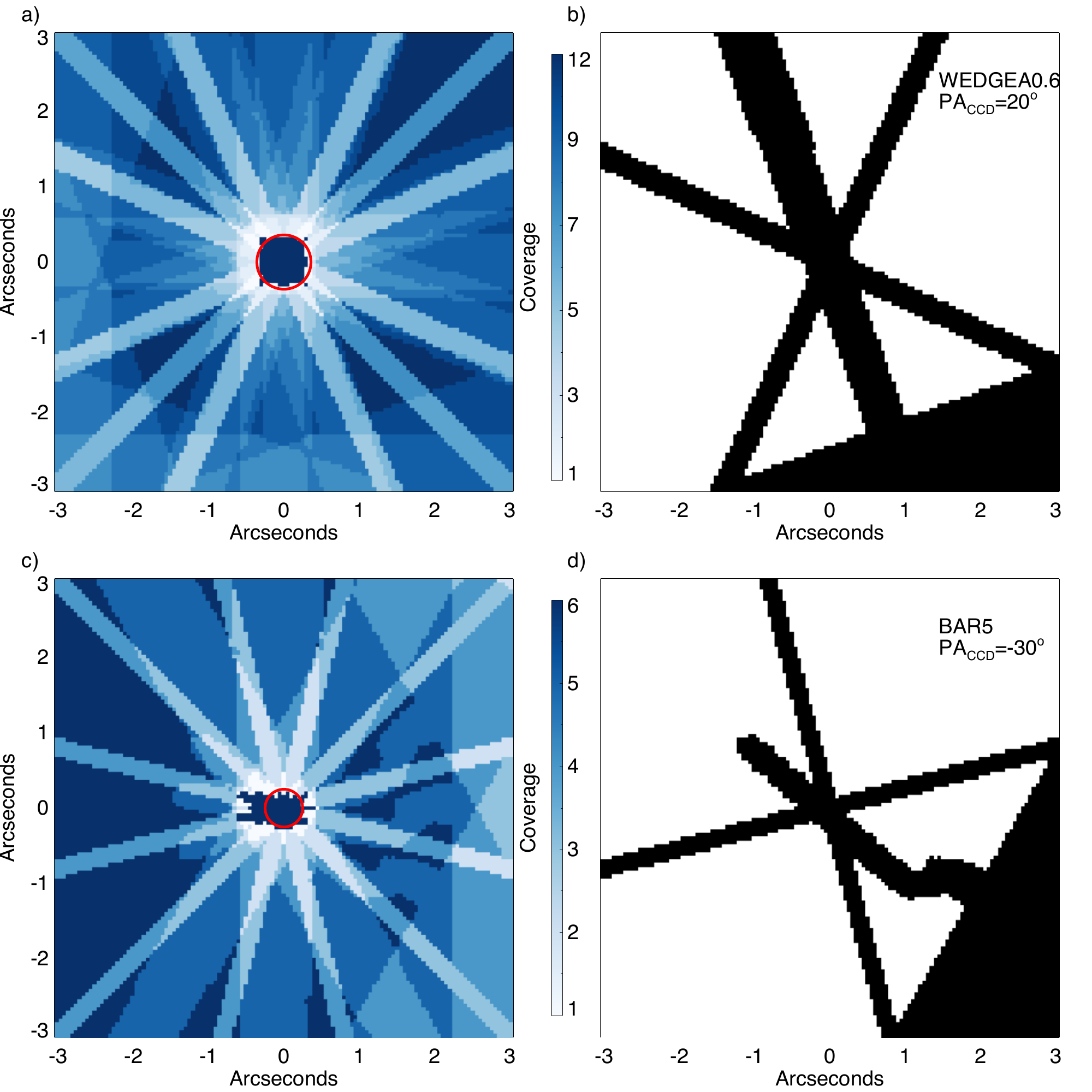}  
\end{center}
\caption 
[Examples of angular coverage gain with the STIS 50CORON mask.]{ \label{fig:f2} Examples of angular coverage gain with the STIS 50CORON mask. Panel a) demonstrates a hypothetical observing program that achieves full angular coverage of a target with a combination of the WEDGEA0.6 and WEDGEA1.0 masks using six separate spacecraft orientations as described in the text. The red circle denotes an inner working angle of $0\farcs45$. Panel b) shows an example mask from one spacecraft orientation that demonstrates regions where there is no data due to the diffraction spikes of the star and the occulter. Panel c) demonstrates a hypothetical observing program that combines BAR5 and WEDGEA1.0, obtaining full angular coverage beyond $0\farcs5$, and an inner working angle of $0\farcs25$ with a total of 3 {\it HST} orbits as described in the text. Panel d) shows an example orientation as in panel b).} 
\end{figure}

In another example, panel a of Figure \ref{fig:f2} shows the coverage in an image with a hypothetical {\it HST} program that utilizes a sequence of 6 relative spacecraft orientations ($-20^\circ, 0^\circ, +20^\circ, +70^\circ, +90^\circ, +110^\circ$) in six separate orbits to maximize angular coverage with an inner working angle of $0\farcs45$ from a combination of shallow images at WEDGEA0.6 and deeper images at WEDGEA1.0. This strategy, including two additional orbits for a reference PSF star, has been used to characterize debris disks around a variety of stars \cite{schneider14}. One scheduling consideration for this approach is the fact that allowed spacecraft orientations are different through the year and restricted---thus limiting opportunities to realize six significantly different orientations within a single Cycle or year.  Often, a gap in observations is required between two sets of orientations. This gap can be 3--6 months, depending on the location of the target on the sky.  Figure \ref{fig:f22} demonstrates a roll angle report of HD 38393 generated with version 26.1 of the Astronomer's Proposal Tool (APT), which can aid in scheduling specific orientations on the sky for a given scene and target, since it lists all possible scheduling opportunities without regard to additional constraints caused by other programs or the availability of HST guide stars.

An approach that obtains a similar inner working angle with fewer orbits is to combine the BAR5 and WEDGEA1.0 aperture locations as demonstrated in recent images of TW~Hya as part of Program 13753.  In this case, nearly $360^\circ$ coverage down to $0\farcs5$ was achieved with three spacecraft orientations ($0,\pm20^\circ$), and partial angular coverage down to an inner working angle of $0\farcs25$\cite{debes17}. The coverage map for three visits of that program is shown in panel b of Figure \ref{fig:f2}. For edge-on circumstellar disks, a combination of BAR5 and WEDGEB1.0 may be a preferable choice to obtain shallow images of circumstellar material close to the star and deeper images further from the star with a minimum of spacecraft orientations, since these two aperture locations are parallel to each other.

Another approach that achieves nearly $360^\circ$ coverage with only two orientations at small inner working angles with short exposure times is to combine BAR5 with lower left and lower right locations of BAR10. While BAR5 is a fully supported aperture position with STIS which can be selected in APT, the lower left and lower right positions of BAR10 can be utilized by executing a POS-TARG, or small angle maneuver from the BAR10 central position. The BAR10 positions have been demonstrated to be moderately effective \cite{schneider17}, but have not yet been combined with BAR5 for any existing programs. However, by combining the 3 positions, one could obtain in principle 360$^\circ$ coverage to an inner working angle of $0\farcs2$, as demonstrated in Figure \ref{fig:f2}.  One potential risk to this approach is that flatfield images of the 50CORON aperture indicate up to a $3\%$ decrease in sensitivity close to all wedge positions, but it is particularly noticeable near BAR10, where such sensitivity decreases can extend up to $10$--$20$ pixel from the BAR10 edges.  Currently, these sensitivity deviations are not corrected within the {\sc calSTIS} pipeline\cite{hodge98}. Users are encouraged to download coronagraphic flatfield images from the archive if their scientific program requires a high degree of photometric repeatability. In most cases, PSF wing intensity non-repeatability likely will dominate over these effects.

At least one program has also experimented with placing targets via a POS-TARG at the edge of the BAR5 aperture (i.e., Program 13725, PI: P.~Kalas). This will trade angular coverage in a single exposure for currently unknown performance of the edge of BAR5---since the occulting finger is bent, it is currently unclear what impact that may have on photometric accuracy or repeatability of contrast, compared to the nominal BAR5 position that has been investigated.

\begin{figure}
\begin{center}
\includegraphics[width=\textwidth]{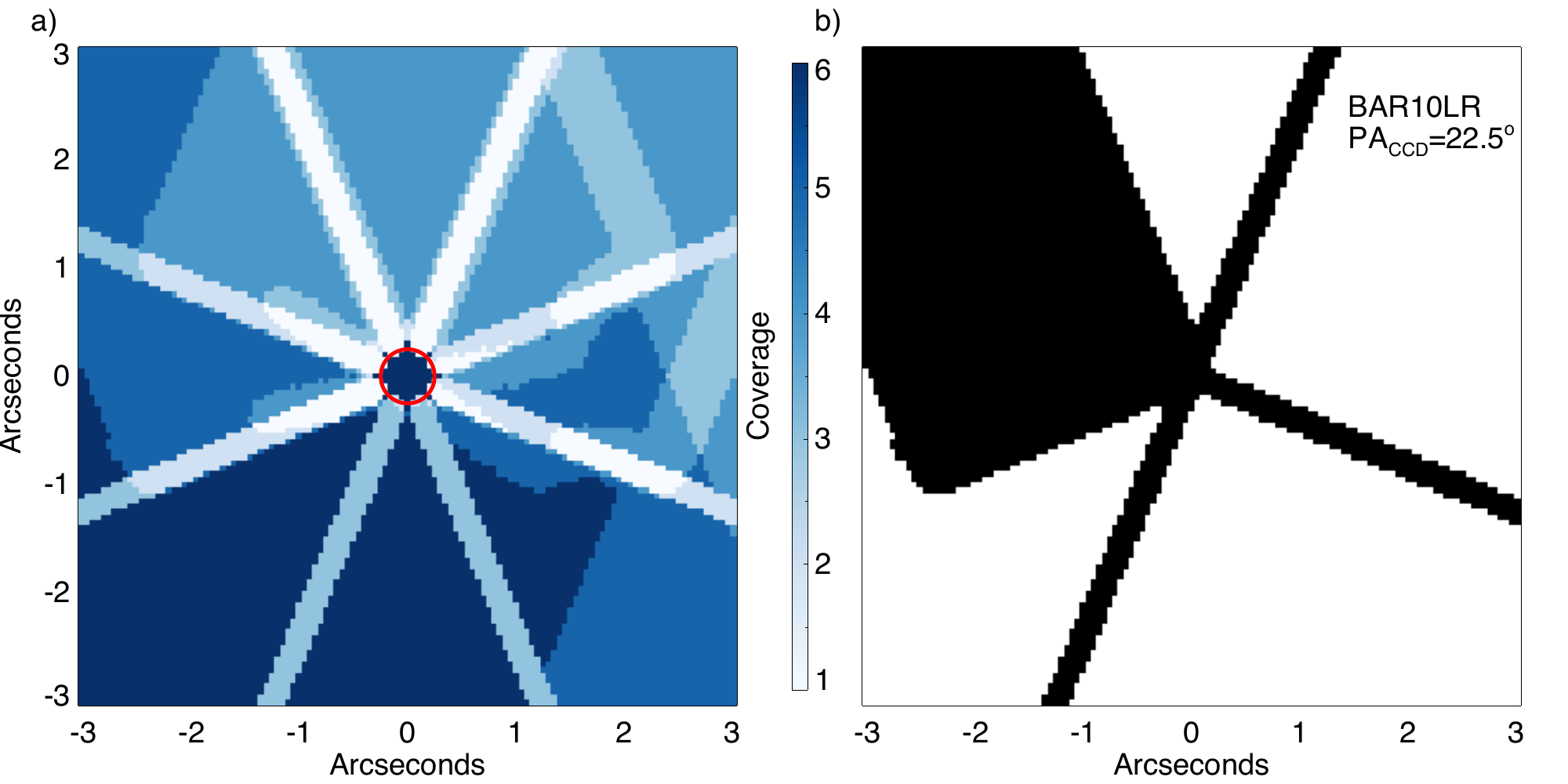}  
\end{center}
\caption 
[Additional example of angular coverage gain with the STIS 50CORON mask.]{ \label{fig:f2prime} Similar to Figure \ref{fig:f2} but for the hypothetical observing program that demonstrates the most efficient way to obtain full angular coverage down to ${\sim}0\farcs2$, using the lower corners of the BAR10 aperture location and BAR5, which is described more fully in the text. Panel a) shows the coverage assuming two spacecraft orientations with the three aperture locations in each orientation, and panel b) shows an example mask for one orientation.} 
\end{figure}

\subsection{Strategies that Reach Highest Contrast}

Several strategies can help reach higher contrast achieved with STIS coronagraphy. These include strategies for selecting the best reference stars, and careful selection of exposure times to maximize the total counts in the PSF without driving the region of interest into saturation. Additionally, we review here the sources of noise that are present in coronagraphic observations, including noise terms that are generally not accounted for within the traditional signal-to-noise ratio (SNR) calculations that are performed for the direct imaging of sources. 

If the science program requires a reference star, matching the color of the reference star across the STIS bandpass is critical for ensuring the best contrast results. Details of recommended color matches are given in Ref.~\citenum{grady05}, but we review here---typically $B-V$ colors must be within 0.05, with closer color matches and spectral type matches working the best. Great care should be taken in considering the SED of the target and reference over the full bandpass of STIS, which runs from ${\sim}0.20$ $\mu$m to ${\sim}$1.03~$\mu$m (Section~\ref{sec:hi-c}), suggesting that better stellar matches can be made if one considers the full spectral energy distribution of an object in that range, or at the very least between $B-V$ and $V-I$ if more complete photometry is not available. This is especially true with very blue or very red stars, where the effective bandpass of the PSF is skewed to wavelengths shorter or longer than $\sim$0.58~$\mu$m. Interestingly, the broad sensitivity also means that STIS is {\em more} sensitive to substellar objects than a typical visible light imager, since it retains sensitivity to photons from 0.8-1.0~$\mu$m, where L and T dwarfs still retain significant flux.

In the case of reference star selection, an ideal candidate is reasonably close in the sky to the target, usually within $10^\circ$. This minimizes the thermal changes in the observatory between the target star and the reference when observed in contiguous orbits without other intervening telescope repointings. If multiple orbits on the target are executed, it is best to have the reference star in either the second or third orbit in the sequence, since most of the thermal settling of the telescope can occur within the first orbit after a slew. Large thermal swings can drive defocus from differential heating across the support structure, which is the cause of telescope ``breathing''

Table \ref{tab:t1} lists observed count rates in the PSF wings of a selection of stellar spectral types in units of $e^{-}$~s$^{-1}$ and using the BAR5 aperture location. The count rates were determined by taking azimuthally averaged surface brightness profiles of observed STIS PSF wings while avoiding diffraction spikes, measuring the median count rate at a given radius, and multiplying by the detector gain. Calculations of the count rate in the peak pixel come from the STIS Exposure Time Calculator (ETC)\footnote{\url{http://etc.stsci.edu/etc/input/stis/imaging/}}. Note that the maximum count rate close to the mask edge can vary from this median value by as much as 20\% compared to the median value, users should be aware of this if they do not wish to saturate directly at the mask edge, especially if they are executing small dithers behind a mask aperture position. Table \ref{tab:t1} provides a handy rule of thumb for the maximally allowed exposure time for a given BAR or WEDGE position assuming a target of 80\% full well and a saturation limit of 120,000 $e^{-}$~s$^{-1}$ for gain $G = 4$. Previous observations with STIS show that the actual full well varies somewhat across the detector \cite{proffitt15}, and is thus a consideration when doing a more careful calculation for how long to expose a target. The 20\% margin provides safety for sub-pixel dithers behind the aperture position and/or miscentering of the target.

\begin{figure}
\begin{center}
\includegraphics[width=1.0\textwidth]{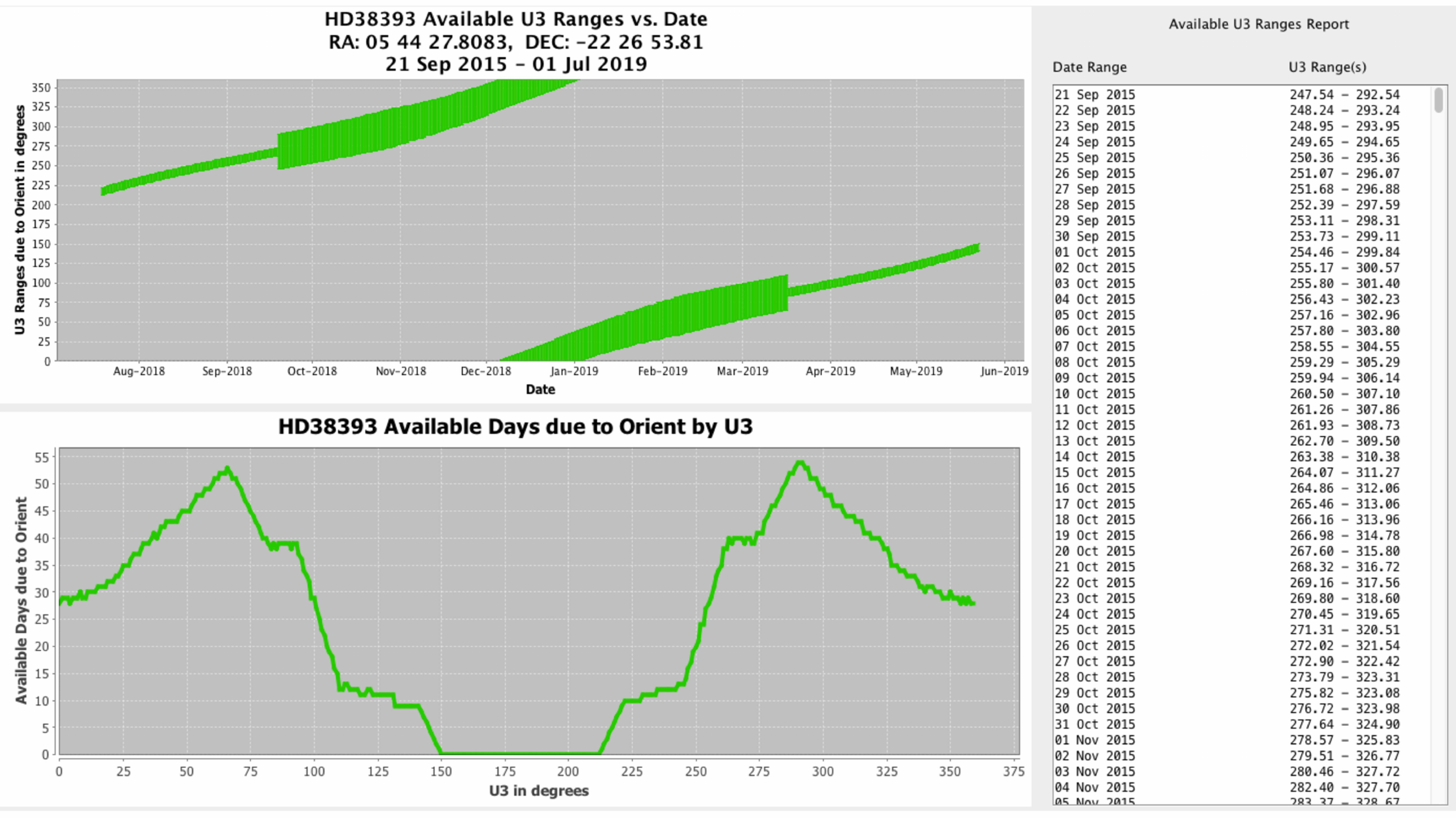}  
\end{center}
\caption 
[Example roll range report for STIS coronagraphic observations of HD~38393.]{ \label{fig:f22} Example roll range report for STIS coronagraphic observations of HD~38393, generated by version 26.1 of the Astronomer's Proposal Tool. The U3 angle ranges correspond to the ``ORIENT'' constraint in APT and for STIS, sky PA $=$ U3$-45^{\circ}$. These reports provide guidance to observers for scheduling their targets with a given orientation. These reports do not guarantee that targets will be observed at these times, as the {\it Hubble Space Telescope}'s observing schedule is highly constrained. Often observers must relax orientation constraints to provide more observing opportunities.
} 
\end{figure}

\begin{table}
\caption[Observed count rates as a function of angular separation.]{\label{tab:t1} The observed count rates as a function of radius for the stellar halo of different spectral types, avoiding diffraction spikes and occulters. S$_{x}$ are count rates in the PSF wings at a $x\arcsec$.}
\begin{center}
\begin{tabular}{cccccccc}
\hline
Star & B-V & V mag & Spectral Type & S$_{0}$ & S$_{0.2}$ & S$_{0.5}$ & S$_{1.0}$ \\
& & & & (e$^{-}$ s$^{-1}$) & (e$^{-}$ s$^{-1}$) & (e$^{-}$ s$^{-1}$) & (e$^{-}$ s$^{-1}$)\\
\hline
$\delta$ Dor & 0.21 & 4.36 & A7V & 1.05$\times$10$^{8}$ & 3.1$\times$10$^4$ & 1.8$\times$10$^4$ & 4.5$\times$10$^3$ \\
HD 38393 & 0.47 & 3.60 &  F6V & 2.13$\times$10$^{8}$ & 8.5$\times$10$^4$ & 6.1$\times$10$^4$ & 9.9$\times$10$^3$ \\
HD191849 & 1.45 & 7.96 & M0V & 5.02$\times$10$^{6}$ & 1.8$\times$10$^3$ &  1.3$\times$10$^3$ & 2.0$\times$10$^2$ \\
\hline
\end{tabular}
\end{center}
\end{table}

Another consideration is the selection of an exposure time for individual read outs of the CCD. The upper limit to the photon limited contrast possible in a single orbit is dictated by the total exposure time, and the efficiency with which a readout can be obtained. The typical readout time of the STIS CCD is roughly 1 minute, but can be as small as 20 seconds if subarrays with a small number of rows are used. Overheads thus limit the total counts in the PSF per orbit under the restriction of not saturating for a given mask position. If we assume that we design a program that has counts close to the full well at the edge of the BAR5 mask position in each exposure ($T_{\rm exp}$), we can calculate the total integration time available for observing a star within a typical 50-minute {\it HST} orbital visibility window ($T_{\rm int}$), accounting for readout and data management overheads and assuming a $1024\times90$ pixel sub-array with a readout time of 20~s. We select this particular sub-array since it represents a compromise between quick read-out time and spatial coverage. The total exposure times and counts in the PSF per orbit are presented in Table \ref{tab:counts}. Total exposure times were calculated using version 26.1 of the Astronomer's Proposal Tool in the Phase II proposal mode, and the count rates were scaled from Table \ref{tab:t1}. Users should adopt these values as a rough rule of thumb, and do their own more detailed calculations for their specific program requirements. 

Table \ref{tab:counts} demonstrates that STIS will obtain similar number of total counts within the PSF wings (to within 50\%) within 1 orbit and with the gain ($G$) of the CCD set to 4 for stars with $9.5<V<3.5$. $G$=4 is in general preferred for high contrast imaging due to the fact that more counts can be collected within a pixel before it saturates. The range of stellar magnitudes is not dissimilar to the range of magnitudes that are effective with high order adaptive optics coronagraphs such as {\it VLT}/SPHERE\cite{fusco06}, or the Gemini Planet Imager\cite{macintosh06}. It should be noted that for very bright stars high contrast imaging is incredibly inefficient due to the overhead associated with sub-array readout times and buffer dumps--for a 50 minute visibility window only 19.2~s in total is spent on source for a V=3.5 star. One significant difference to the ground is that STIS' Strehl ratio does not degrade for fainter sources. Thus for objects with $V>10$, STIS' contrast performance is dictated by the total exposure time available to expose the wings of the PSF relative to the background limited sensitivity of the exposures.

 \subsection{Calculating STIS Raw Coronagraphic Performance (without PSF-Subtraction)}
 \label{sec:performance}
Currently the STIS ETC does not include a straightforward way to estimate the raw performance of STIS coronagraphic modes for a given exposure time and target star. For STIS imaging modes, the ETC does include noise sources such as solar system zodiacal light, CCD dark current (which has slowly been growing with time due to its exposure to the harsh radiation of space), and CCD read noise. Light from the occulted star's PSF wings, however, is by far the largest contributor in particular at small stellocentric angles, but it is currently not considered in the ETC. Further, stochastic noise from quasi-static aberrations and jitter are currently not considered. A simple analytical case for noise sources can be assumed following the methodology of a typical background limited observation. Using information provided in the {\it Hubble Space Telescope} ETC User Manual\footnote{\url{http://etc.stsci.edu/etcstatic/users_guide/1_2_1_snr.html}}, we first define the noise coming from the STIS CCD detector, $\sigma_{\rm det}$:

\begin{equation}
\label{eq:e1}
\sigma_{\rm det}^2=N_{\rm pix}t_{\rm int} \left(S_{\rm dark}+\frac{\sigma^2_{\rm RN}}{t_{\rm exp}}\right),
\end{equation}
where $N_{\rm pix}$ is the number of pixels in a photometric aperture, $t_{\rm int}$ is the total integration time of an observation set (i.e., $t_{\rm int}=Nt_{\rm exp}$), $S_{\rm dark}$ is the dark rate in units of e$^{-}$ s$^{-1}$ pixel$^{-1}$, $\frac{\sigma_{\rm RN}^2}{t_{\rm exp}}$ is the read noise in each individual exposure (also known as CR-SPLITs in STIS terminology. For STIS, $\sigma_{\rm RN}=8.2$ e$^{-}$ pixel$^{-1}$ exposure$^{-1}$ for gain $G = 4$. We ignore the noise induced from cosmic ray hits.

We then define the important background noise sources for space observatories, including light from the Solar System zodiacal dust disk ($S_{\rm zodi}$) and scattered light from Earthshine ($S_{\rm Earthshine}$), also in units of e$^-$ s$^{-1}$ pixel$^{-1}$:

\begin{equation}
\label{eq:e2}
\sigma_{\rm bkg}^2=N_{\rm pix}t_{\rm int} \left(S_{\rm zodi}+S_{\rm Earthshine}\right).
\end{equation}
We assume the expected detector noise properties of STIS projected to mid-Cycle 26 for the middle of the detector and with gain $G = 4$ ($S_{\rm dark}=0.025$). In practice, Earth shine and zodiacal light are much smaller than other noise sources. 

For high contrast imaging, one also needs to take into account the photon noise from the PSF wing intensity ($S_{\rm PSF}$), as well as the ``speckles'' from both static aberrations as well as slowly varying wavefront errors ($S_{\rm spec}$). Static speckles can typically be removed via subtraction of a reference star. Slowly varying speckles correlate with the thermal and focus state of the telescope, the so-called telescope breathing. For STIS, because of its wide bandpass, the speckles are actually spoke-like in shape and extend radially (See Figure 8 of Ref. \citenum{grady03}). The average observed intensity $I$ is comprised of the static, aberrated coronagraphic PSF intensity $I_{\rm C}$ as well as speckles from high frequency telescope jitter $I_{\rm J}$ and speckles from quasi-static errors $I_{\rm QS}$\cite{soummer07,pueyo18}. The photon noise from these three contributions is given by

\begin{equation}
\label{eq:e3}
\sigma_{\rm P}^2=N_{\rm pix}t_{\rm int} \left(I_{\rm C}+ I_{\rm J} + I_{\rm QS}\right).
\end{equation}

$I_{\rm C}$ can be determined from a log-log interpolation of Table \ref{tab:counts}, or by referring to the publicly available, azimuthally averaged PSF intensities provided on the STIS Instrument website\footnote{\url{http://www.stsci.edu/~STIS/coronagraphic_bars/GO12923raw.dat}} for a more detailed profile. These intensities are calculated relative to the peak pixel of the STIS PSF and are per pixel intensities. The other intensities are assumed to be much less than I$_{\rm C}$ and are ignored.

\begin{table}
\caption[Counts per pixel as a function of angular separation.]{\label{tab:counts} Effective number of counts as a function of radius within a typical {\it Hubble} orbital visibilty of 50 minutes and assuming typical overheads resulting in a total integration time $T_{\rm int}$. We assume observations are taken in gain $G=4$ mode. Observers should determine the specific exposure times obtained for a given target and its visibility window.}
\begin{center}
\begin{tabular}{cccccc}
\hline
V & T$_{\rm exp}$ & T$_{\rm int}$ & Counts$_{0.2}$ & Counts$_{0.5}$ & Counts$_{1.0}$ \\
   &    (s)                     &  (s)  & pixel$^{-1}$ & pixel$^{-1}$ & pixel$^{-1}$ \\
\hline
3.5 &  0.2 & 19.2 & 6.6$\times$10$^5$ & 2.3$\times$10$^5$ & 3.4$\times$10$^4$ \\
4.5 & 0.5 & 36 & 5.0$\times$10$^5$ & 1.7$\times$10$^5$ & 2.5$\times$10$^4$ \\
5.5 & 1.2 & 117.5 & 6.5$\times$10$^5$ & 2.2$\times$10$^5$ & 3.3$\times$10$^4$ \\
6.5 & 3.1 & 268.8 & 5.9$\times$10$^5$ & 2.0$\times$10$^5$ & 3.0$\times$10$^4$ \\
7.5 & 7.8 & 566 & 4.9$\times$10$^5$ & 1.7$\times$10$^5$ & 2.5$\times$10$^4$ \\
8.5 & 19.5 & 1014 & 3.5$\times$10$^5$ & 1.2$\times$10$^5$ & 1.8$\times$10$^4$ \\
9.5 & 50 & 1600 & 2.2$\times$10$^5$ & 7.6$\times$10$^4$ & 1.1$\times$10$^4$ \\
10.5 & 126 & 2142 & 1.2 $\times$10$^5$ &  4.1$\times$10$^4$ & 6.0$\times$10$^3$ \\
\hline
\end{tabular}
\end{center}
\end{table}

We finally construct the total noise and put it into units of Analog to Digital Units (ADU: i.e., the counts within a STIS calibrated exposure obtained from the MAST archive) via multiplication of the gain ($G$) of the CCD, which is typically either 1 or 4. We also account for reference star subtraction with the intensity scale factor $A_{\rm ref}$, which corresponds to ($1+F_\star/F_{\rm ref}$) where $F_\star/F_{\rm ref}$ is the ratio of the flux from the target and reference star observations, and $A_{\rm t}$, which corresponds to ($1+t_{\rm exp,\star}/t_{\rm exp, ref})$
\begin{equation}
\label{eq:e4}
\sigma_{\rm tot}=G^{-1} \sqrt{A_{\rm ref} \left(\sigma_{\rm spec}^2+\sigma_{\rm P}^2\right)+ A_{\rm t} \left(\sigma_{\rm bkg}^2+\sigma_{\rm det}^2\right)} .
\end{equation}

\subsection{Estimating Systematic Speckle behavior in STIS high contrast images}
\label{sec:speckle}
It is important to note that $\sigma^2_{\rm spec}$ has not formally been studied with STIS. Ground-based AO systems typically have speckle intensities that follow a modified Rician probability distribution function (PDF) or a Weibull PDF, rather than a Gaussian PDF of the mean intensity \cite{marois08,soummer07,stangalini17}. Recent work has also shown that with extreme-AO fed systems, atmospheric decorrelations and internal optics decorrelations grow exponentially within the first tens of seconds and then grow linearly on the timescale of minutes to hours \cite{milli16}. Atmospheric driven speckles have intensity distributions as a function of angular separation that mimic a Moffat profile halo with a full-width half max that is roughly proportional to the characteristic length of the atmospheric turbulence \cite{racine99}. We expect that the majority of speckles in STIS will be generated by a combination of jitter, which operates on short timescales, and focus changes, which operate on orbital visibility timescales ($\approx$50~min).

In this case \cite{soummer07,pueyo18}, the noise associated with the speckles can be directly related to the speckle intensities of each type as well as their cross-terms, as well as the total number of independent speckle realizations, $N=t_{\rm int}/t_{\rm exp}$:

\begin{equation}
    \label{eq:eq5}
    \sigma_{spec}^2=N_{\rm pix}^2 N  F_\star^2 \left[I_{\rm J}^2+NI_{\rm QS}^2+2 I_{\rm C}\left(I_{\rm J}+NI_{\rm QS}\right)+I_{\rm J}I_{\rm QS}\right]
\end{equation}

Directly quanitfying each term in Equation \ref{eq:eq5} requires careful analysis and assumes that these terms have well-defined timescales. Since most coronagraphic exposures are taken with $t_{\rm exp}$ of a few seconds to a few hundred seconds, one can estimate $\sigma_{\rm spec}$ by calculating noise in each pixel and measuring how much it exceeds that expected from the PSF wings alone. Such an analysis by definition ignores the possible spatial correlations due to the spoke-like shape of STIS speckles. Accounting for spatial correlations is beyond the scope of this paper, but we assume that one cannot average over pixels in a small photometric aperture due to this correlation. 

We investigated the measured variance of speckles in several STIS coronagraphic images using a combination of the BAR5, WEDGEA0.6, and WEDGEA1.0 aperture locations. In particular, we studied the intensity RMS in the HD 38393 ($V=3.6$) observations with BAR5 from Program 14426 (PI: J.~Debes), and HD~30447 ($V=7.9$), HD~141943 ($V=8.0$), HD~35841 ($V=8.9$), and HD~191089 ($V=7.2$) from Program 13381 (PI: M.~Perrin); therefore, our observations sample the state of the telescope between 2013 and 2015. For the Program 13381 targets, we investigated the noise in both WEDGEA1.0 and WEDGEA0.6 aperture locations (See Table \ref{tab:noise}. Since each of these targets had multiple exposures of the CCD within a single orbit in at least two orbits, we first calculated a root mean square (RMS) noise per pixel as a function of time for the images of one star. We azimuthally averaged the resulting 2-D RMS over time to construct a radial distribution of RMS. To determine the contribution from speckle noise we made a few assumptions. Firstly, we assumed that the majority of the noise observed is due to deterministic photon statistics as estimated from the STIS CCD detector parameters, the brightness of the star, and the observed average PSF wing intensity with static aberrations as determined by Equation~\eqref{eq:e3} and assuming that the speckle noise is 0. At large angular distances for short exposure times, we see excellent agreement with these predictions. We also assumed that the speckle noise in one detector pixel is given by $\sigma^2_{\rm spec}\propto\left[F_\star f(\theta)\right]^2t_{\rm exp}^2$, where $F_\star$ is the flux from the central astrophysical object in the peak pixel in units of e$^{-}$ pixel$^{-1}$, and $f(\theta)$ is the speckle RMS dependence as a function of the angular separation to the central object. This functional form for the speckle variance as function of angular distance from the star is roughly equivalent to the normalized speckle variance discussed in equation 36 of \cite{soummer07}. This means that the variance we measured should scale both with exposure time, and with the brightness of the central object.

\begin{table}
\caption[Stars and wedges used for systematic speckle noise calculation.]{\label{tab:noise} Stars used for calculation systematic speckle noise at BAR5, WEDGEA0.6, and WEDGEA1.0.}
\begin{center}
\begin{tabular}{lcccc}
\hline
Star & $V$ & $F_{\rm \star}$ & Aperture Location & $t_{\rm exp}$ \\
     &   & (electrons) & & (s) \\
\hline
HD 38393 & 3.6 & 2.09$\times$10$^8$ & BAR5 & 0.2 \\
\hline
HD 30447 & 7.9 & 4.14$\times$10$^6$ & WEDGEA0.6 & 60 \\
         &     & & WEDGEA1.0 & 540 \\
\hline
HD 141943 & 8.0 &  3.82$\times$10$^6$ & WEDGEA0.6 & 60 \\
          &     & & WEDGEA1.0 & 567 \\
\hline
HD 35841 & 8.9 & 1.59$\times$10$^6$ & WEDGEA0.6 & 120 \\
         &     &  & WEDGEA1.0 & 485 \\
\hline
HD 191809 & 7.2 & 7.72$\times$10$^6$ & WEDGEA0.6 & 32 \\
          &     & & WEDGEA1.0 & 483 \\
\hline
\end{tabular}
\end{center}
\end{table}

\begin{figure}
\begin{center}
\includegraphics[width=0.6\textwidth]{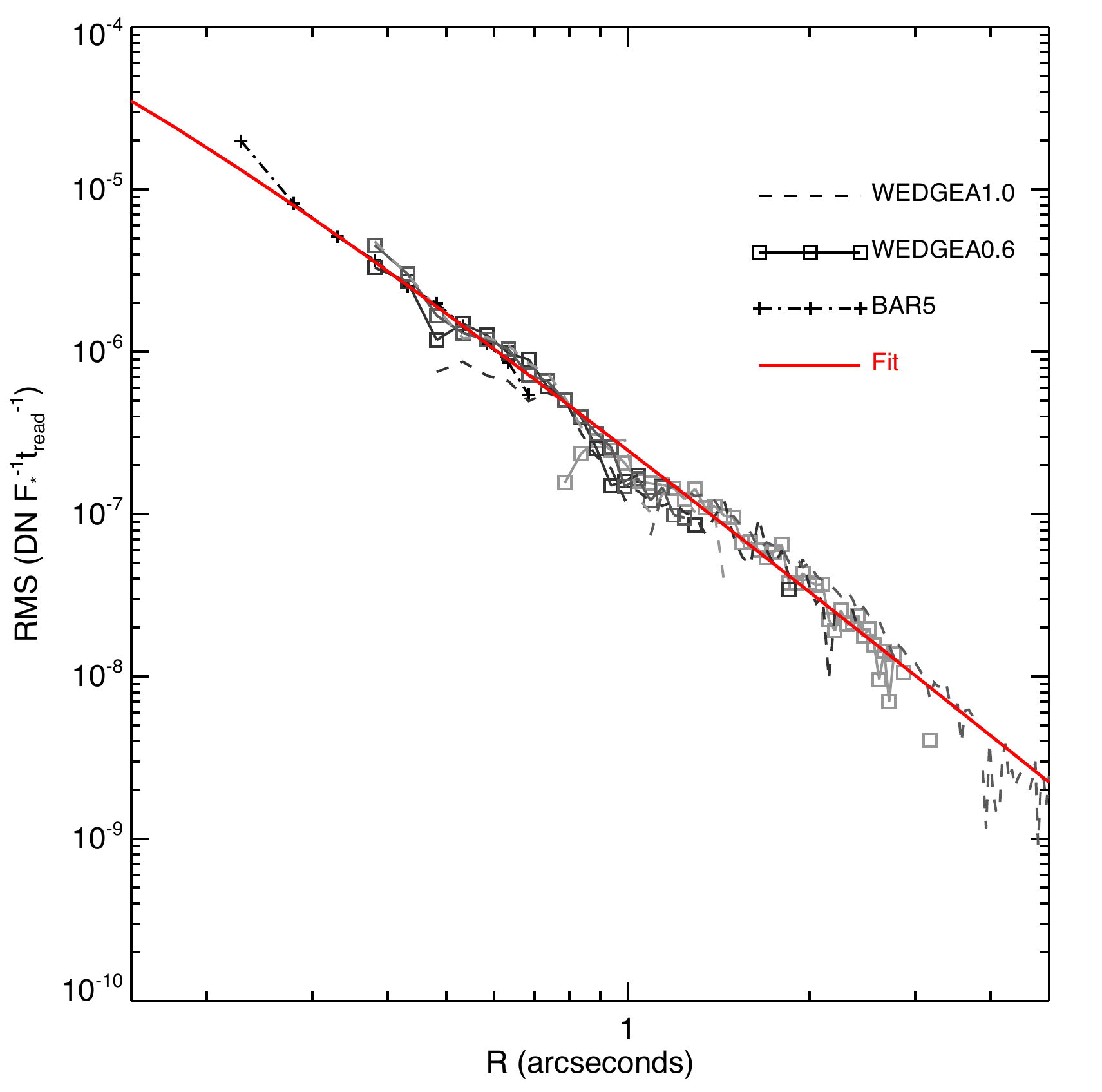}  
\end{center}
\caption 
[Measured STIS speckle RMS as a function of radius from a host star.]{ \label{fig:f2aa} Measured STIS speckle RMS as a function of radius from a host star. We have calculated the residual speckle RMS from multiple coronagraphic observations for various stars after normalizing by stellar flux and exposure time and by subtracting in quadrature the expected RMS from photon noise of the stellar PSF wings. We fit the combined observations into a single function that describes the dependence of the RMS (assumed to be equivalent to the speckle variance) as a function of distance from the star.
} 
\end{figure}

Taking our RMS measures per star, we calculated the residual noise above the predicted photon and detector noise, especially in regions where the noise was much larger than the predicted variance. We then scaled the residual noise by both the exposure time (which ranged from 0.2s to 540s) and the expected flux in electrons for each star as calculated by {\sc Psynphot} for the relevant apparent V magnitude and spectral type of each target (the stellar spectral types were F3--G2). Figure \ref{fig:f2aa} demonstrates our resulting empirical measure of $f(\theta)$ for our targeted stars, with the exception of HD~141943. In the case of HD~141943, the observed noise closely matched the predicted noise from just the PSF wings and the detector, precluding an accurate fit of additional systematic residuals and implying, for those observations, the speckle noise was a factor of two lower than compared to the other observations. We thus expect our results for the majority of the stars are ``typical'' and that while some variation exists due to the state of the telescope the variability doesn't exceed a factor of two for nominal telescope operations for that time period. More detailed estimates on the distribution of speckle noise levels and what they might depend on as a function of the STIS lifetime is left for future work. We recommend that anyone using this approach to predict performance to take coronagraphic observations from the MAST archive as close in time as possible to verify the latest speckle behavior of STIS and to build some margin, if possible, into their observational design. Figure \ref{fig:f2aa} also demonstrates that our assumption on the dependence on stellar flux and exposure time were correct over a fairly typical range of stellar brightnesses and exposure times relevant for typical STIS observations, and that the primary dependence on angular separation, $f(\theta)$ can be fit empirically. By averaging together all the existing residual noise curves, we fit a function to the angular dependence of the speckle RMS with a form of $\sigma=C(r/r_{o})^{\alpha}$, where $C=1.3\pm0.2\times10^{-3}\pm F_\star$, $\alpha$=-2.86$\pm$0.06, and $r$ is in units of milliarcseconds. The constant $r_{o}$ is equal to the platescale of STIS, 50.7~mas. The uncertainty of each noise measure was typically between 20-50\% across the four stars. The dependence on angular radius is similar to the dependence of sensitivity limits derived in deep STIS coronagraphic images of HD~207129\cite{schneider16}. Putting this together, $\sigma_{\rm spec}$ for our noise model would become:

\begin{equation}
\label{eq:e6}
\sigma_{\rm spec}^2 = N_{\rm pix}^2 \frac{t_{\rm int}}{t_{\rm exp}}\left(F_\star f(\theta)\right)^2
\end{equation}
This encapsulates our assumptions about how the speckle variance scales with exposure time, and implicitly assumes that speckles are likely spatially correlated on small angular scales. This is appropriate given the fact that a single STIS CCD pixel is close to $\lambda_c/D$, where $\lambda_c$ is the central wavelength of of the CCD bandpass and multiple pixels will encompass multiple speckles. Additionally, since the sensitivity of the CCD spans from 2000-10000\AA, individual speckles can span a large number of STIS pixels in the radial direction. As an example, if an aberration caused a speckle at a radius of 4$\lambda$/D, it would span from 68~mas to 343~mas, or the equivalent of 5~pixels in radial extent.

We now include Equation~\eqref{eq:e6} in Equation~\eqref{eq:e3} to obtain the final estimate of the noise. In the next subsection we compare this to a suite of observations from Program 14426.

The physical origin of this variance profile, which roughly tracks the power-law decrease of the STIS PSF wings is likely caused by speckle pinning\cite{soummer07}. Since the STIS wedges and bars do not optimally suppress the diffraction pattern, STIS coronagraphic images approximate direct imaging at a high strehl ratio. 

We note that a similar procedure could be implemented during the {\it James Webb Space Telescope} ({\it JWST}) commissioning in order to quantify speckle statistics for that observatory, as well as for other future high contrast imaging missions. A dedicated calibration program would image sources with increasing exposure times to characterize the raw uncorrected speckle intensity field under differing wavefront control algorithms as well as with reference star subtraction. The performance could then be monitored over the mission lifetime to aid in providing planning information to general observers. It is often a significant stumbling block to successful high contrast imaging if the noise statistics of speckles are not properly accounted for.

An additional subtlety of the speckle noise is in optimizing an observing strategy when one is attempting to use a reference star to obtain high contrast. Using a reference star with a different brightness and exposure time can drastically increase residual speckle noise because it will then be dominated by the speckles for one of the stars. The residual halo described in STIS observations of HR~8799\cite{gerard16} is due in part to the very different exposure times and apparent magnitudes of HR~8799 and its selected reference star.
 

\subsection{Validation of the Coronagraphic Noise Model}
\begin{figure}
\begin{center}
\includegraphics[width=0.6\textwidth]{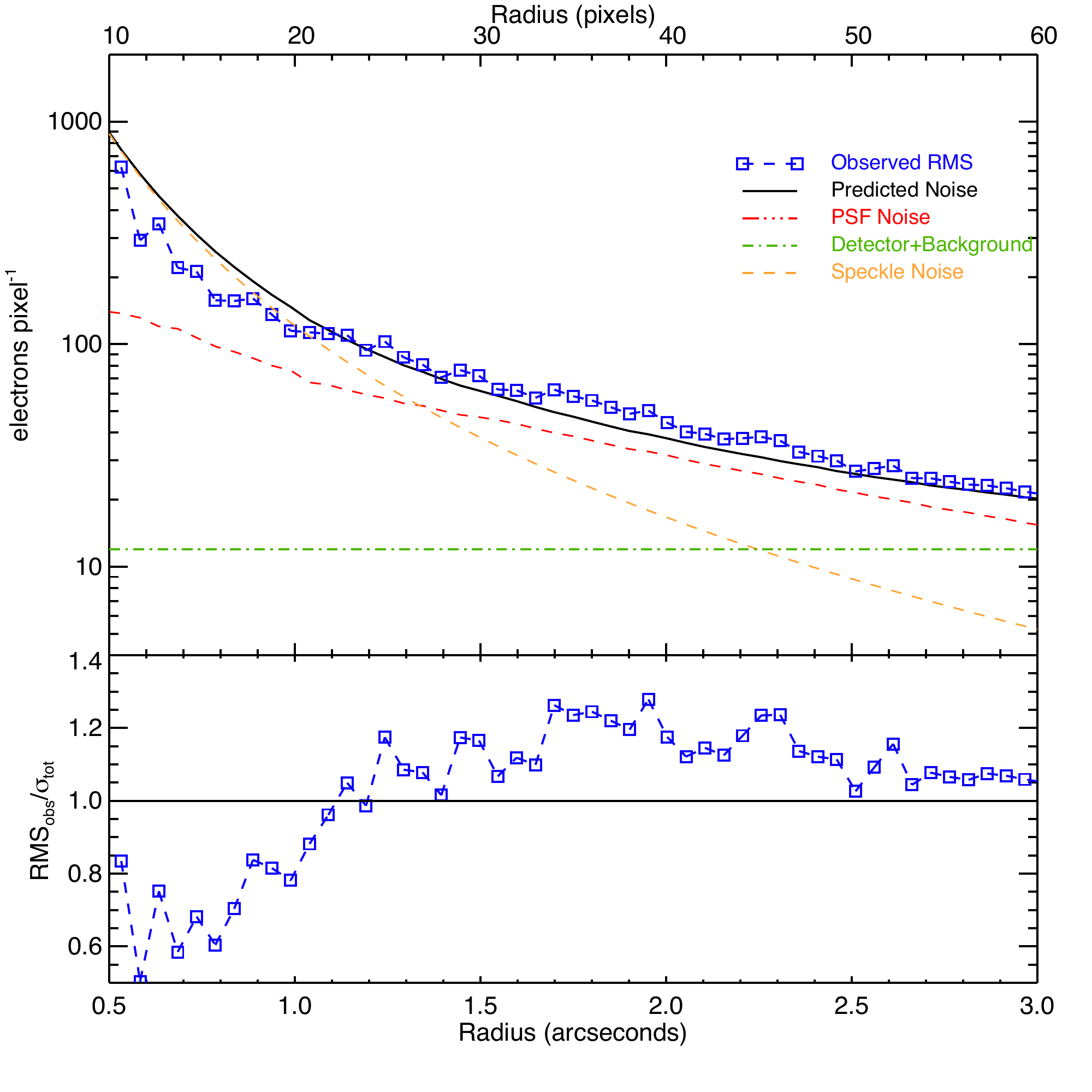}  
\end{center}
\caption 
[Test of the STIS coronagraphic noise model.]{ \label{fig:f4} Test of the STIS coronagraphic noise model. The measured azimuthally averaged RMS as a function of time per pixel for 24 20~s exposures of HR~8799 taken behind the WEDGEA1.0 aperture location compared to predicted noise sources including detector noise (green dashed-dotted line), our empirically derived speckle noise (dashed orange line), photon noise from the PSF wings (red dashed-triple dotted line), as well as the total noise (black solid line). 
} 
\end{figure}

We validate our noise model by comparing the per pixel RMS of 24 exposures of HR~8799 taken as part of Program 12281 (PI: M. Clampin) to predictions. We calculate the RMS over time per pixel as described in Section \ref{sec:speckle}. The exposure time for each CR-SPLIT was 20~s, with 8 CR-SPLITs per exposure. We compared the azimuthally averaged RMS as a function of time by excluding regions containing the WEDGEA1.0 mask and {\it HST} diffraction spikes to the analytical noise formulation given in Equations~\eqref{eq:e1} to \eqref{eq:e4}. Figure \ref{fig:f4} shows the results, along with the individual components of the noise. At this exposure time, the dark and sky backgrounds are negligible compared to the noise in the PSF wings and due to speckles. We selected this dataset because it was taken close in time to the stars for which we measured the speckle noise directly. Looking at the ratio of the radial noise profile relative to the predicted profile, we determine that the median observed noise was 8\% higher than predicted between $0\farcs5-3\farcs0$. Close to the star we overpredicted the expected speckle noise by 40\%. It is important to note that this noise per pixel is not sufficient to predict the contrast sensitivity--one will need a defined photometric aperture as well as an accounting of the degradation of contrast at the inner edge due to incomplete azimuthal coverage due to the occulters and the residual diffraction spikes of the target.

For observation planning, one will also need an estimate of the target source flux and its accompanying photon noise to conduct a full calculation to determine an optimum exposure time. We investigate these complexities further for point sources in Section \ref{sec:pointsource} and Section \ref{sec:disk} for extended sources such as circumstellar disks.

\subsection{The Effects of Charge Transfer Efficiency Degradation and potential impacts to science}

Due to its 21-year lifetime in low Earth orbit since 1997, the STIS CCD has suffered from cumulative radiation damage. The damage primarily takes the form of increasing numbers of hot pixels on the detector, increasing dark rate, and increasing numbers of defects in the silicon substrate that ``trap'' electrons generated from a photon striking the detector. These trapped electrons are then released later than expected on the detector during readout, resulting in an artificial trailing of light away from the readout direction and a degradation in charge transfer efficiency (CTE) \cite{anderson10}.  In the case of a point source on STIS, this will appear as a trail of light that points to the bottom of the image. For an extended source, charge traps will create a low level halo that is asymmetrically stronger a few pixels below the source.

If one is conducting high contrast imaging with STIS, the image is dominated by the stellar PSF wings, and any CTE degradation might show up in the case where the reference PSF has a signifcantly larger or smaller amount of counts in the wings, since degradation is dependent on the total counts in a pixel. The impact of degrading CTE on high contrast imaging is of particular interest to the design and operation of the coronagraphic instrument (CGI) within NASA's proposed {\it WFIRST} mission\cite{nemati16}. While CGI will likely utilize a different type of detector than STIS, the requirements on contrast will be much more stringent.

We investigate the upper limits to CTE degradation impacts with STIS by investigating the behavior of images of the same source taken within the same orbit but with differing exposure times. One example of this is observations of HR~8799 taken in program 12281. Both 20~s and 60~s exposures were obtained with the WEDGEA1.0 aperture position of HR~8799, resulting in a factor of 3 difference in the total counts on the CCD from the PSF wings. Additionally, the program selected a PSF reference star, HIP~117990, that is a factor of 2 fainter than HR~8799. For the PSF exposures, each readout had an exposure time of 318~s, resulting in a factor of 2.6 and 7.8 more counts on the detector than for the 60~s and 20~s HR8799 exposures respectively. 

We constructed images of four cases: a mean combination of $20$~s exposures subtracted by $20$~s exposures from an adjacent orbit (case A), a combination of the PSF images subtracted from the $60$~s exposures (case B), a combination of the $60$~s exposures subtracted from the $20$~s exposures (case C), and a combination of the PSF images subtracted from the $20$~s exposures (case D). In this way we investigate the impact of differing total counts on the detector between reference PSFs with increasing scale factors of $1, 2.6, 3,$ and $7.8$ relative to the target for the four cases. From Case A to Case D, one would expect that the PSF reference star would be less impacted by CTE degradation and one would expect positive residuals preferentially pointing toward the bottom of the detector.

\begin{figure}
\begin{center}
\includegraphics[width=0.6\textwidth]{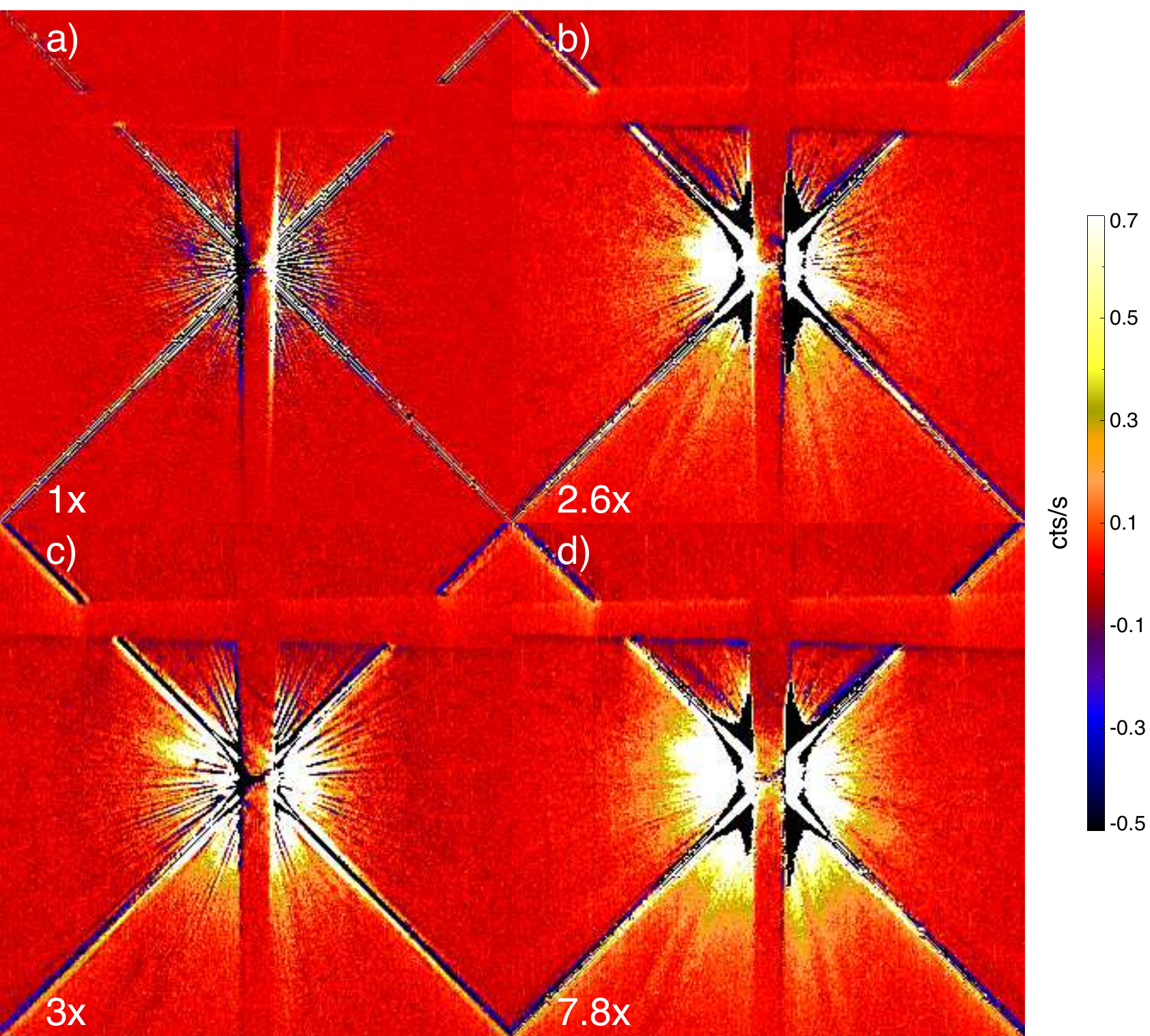}  
\end{center}
\caption 
[The impact of CTE on reference star PSF subtraction.]{ \label{fig:f5} The impact of CTE on reference star PSF subtraction. We investigate the potential impact of CTE on reference star PSF subtraction when the reference star and the target star have differing total counts in an image. Panel a corresponds to the target star HR~8799 subtracted by itself in a later orbit but with the same exposure time. Panel b corresponds to the target star subtracted by a PSF reference that has a factor of 2.6 times more counts on the detector than the target. Panel c corresponds to the target star subtracted by itself where the exposure time of the reference is three times longer. The lower left panel d shows the target star subtracted by a PSF reference that has 7.8 times more counts on the detector than the target. As the difference between target and reference grows, there exists a stronger and stronger residual halo due to a CTE mismatch.
} 
\end{figure}

Figure \ref{fig:f5} shows the resulting subtraction images for each case, demonstrating the presence of increasing residual surface brightness at small stellocentric angular radii as the multiplicative factor of the counts on the detector for the PSF relative to the target increases. While we note an increasing residual halo in the images, this is not an entirely clean experiment since there could be color and focus differences between HR~8799 and its reference (although in this case the color difference between the two objects is relatively small, $|\Delta(B-V)|=0.06$). There appear to be vertical asymmetries in the residuals that are most noticeable along the diffraction spike subtraction residuals and near the bottom of the PSF. At an angular radius 1$\arcsec$, the median surface brightness of the residuals is  $2, 6,$ and $9$ counts s$^{-1}$ pixel$^{-1}$ for cases B, C, and D respectively compared to $1.02\times10^8$ counts s$^{-1}$ estimated to be in the peak pixel of HR~8799. These numbers correspond to residual emission per pixel from CTE degradation to be no more than at a contrast of between $2$ and $9\times10^{-8}$. In the worst cases, where a specifically sharp PSF structure exists the emission can be a factor of two to four brighter (such as at the feature located at the $10$ o'clock position in Figure \ref{fig:f5}.

Based on these results, we recommend that users select PSF reference stars that are close in brightness or slightly brighter and to choose exposure times that ensure the same number of counts on the detector within the PSF wings to better than a factor of 2. This should limit the impact of CTE degradation to a surface brightness that is lower than $10^{-8}$ pixel$^{-1}$ relative to the peak flux of the star at $1\arcsec$ as well as minimize noise from differing levels of speckle noise. These observations will also have the advantage of sampling the breathing within an orbit in a similar way to better match PSF images to the target and minimize subtraction residuals.

The impact of CTE for high contrast imaging has been an area of focus for the CGI of the {\it WFIRST} mission. The CGI will be a technology demonstration of extreme high contrast imaging at small inner working angles that will likely use a photon-counting EMCCD, which will still be vulnerable to CTE degradation due to radiation damage. More work will need to be done to understand whether slight differences in counts on an EMCCD detector could result in similar behavior to the upper limits we place above. Since we do not significantly detect any degradation after more than 20 years in space, it is likely that the largest impact to CGI will be in the loss of signal from faint sources rather than contrast degradation.

\section{High Contrast Imaging at Small Inner Working Angles with BAR5}
\label{sec:push}
The BAR5 aperture location on the 50CORON aperture has a physical half width of $0\farcs15$, and it has been demonstrated to have an effective inner working angle of $0\farcs2$. As the aperture with the smallest inner working angle on 50CORON, BAR5 has achieved reliable high contrast results for bright circumstellar disks \cite{schneider17}. Program 14426 was designed to achieve a working contrast at the BAR5 inner working angle with a goal of obtaining contrasts of $10^{-6}$ or better, comparable in visible-light to near-IR contrasts achieved on the ground.

\subsection{Target Selection}
HD~38393 ($\gamma$ Lep A) is a nearby 1.3 Gyr old\cite{holmberg09} F6V\cite{abt08} star at $d=8.88\pm0.03$~pc\cite{gaia18} in the Ursa Major moving group\cite{king03}, with a common proper motion K dwarf companion ($\gamma$ Lep B) at a projected separation of 95$\arcsec$ (855~au) \cite{hipparcos,mason_wds2001}. It was chosen due to its availability for scheduling, its brightness, and its relative dearth of known companions or circumstellar material. HD~38393 was previously directly imaged with the MMT/CLIO Instrument  \cite{Heinze2010,Vigan2017}. The tabulated contrast sensitivity curve\footnote{\url{http://www.hopewriter.com/Astronomyfiles/Data/SurveyPaper/}} for HD 38393 shows that substellar objects more massive than 15~$M_{\rm Jupiter}$ are ruled out for this star beyond $\sim$13~au. HD~38393 has not been investigated for radial velocity variations \cite{HowardFulton2016}. Due to its proximity to earth, HD~38393 has been the target of  {\em Spitzer} photometric and {\it Keck} nulling interferometer searches for dust, showing a lack of any dust around the system with limits to relative infrared luminosities  $L_{\rm IR}/L_\star<2\times10^{-6}$ for cold dust and $<10^{-5}$ for warmer dust respectively \cite{beichman06,mennessonkin}. 

\subsection{Observing Strategy}
Program 14426 was designed to observe the star HD~38393 over the course of nine orbits\footnote{\label{note:observation}\url{http://www.stsci.edu/cgi-bin/get-proposal-info?id=14426}}. Each orbit was at a different spacecraft orientation, and the orbits were arranged in three sets of three continuous orbits separated by roughly one or two months. The first set of observations executed on 2015 October 21, with the final set of visits executing on 2016 February 26. The final visit was also used to perform a test of the updated Science Instrument Aperture File\cref{note:siaf} position of BAR5 to validate the supported aperture position in APT starting with version 24.1.

HD~38393 was imaged with sub-pixel dithers in a $3{\times}3$ grid behind the BAR5 aperture: on one hand, small mis-centerings on the order of $0.25$ pixel (${\sim}12$ mas) in BAR5 can cause significant differences in the amount of light leaking around the edge of the BAR5 mask, and sub-pixel dithering provides margins against mask and target mis-registration with imperfect initial acquisition, with then improved PSF rejection\cite{schneider17}; on the other hand, it allows for an operational test of a similar dithering strategy proposed for the {\it JWST} coronagraphs \cite{soummer14}. For classical PSF subtraction, sub-pixel dithers can allow better matching of the images between a target star and reference star; while for the statistical post-processing techniques (e.g., LOCI, KLIP, NMF, etc.), it provides additional diversity to the PSF library. This strategy has been suggested to mitigate the tendency of coronagraphic masks on the MIRI and NIRCam instruments to degrade in performance with the expected target acquisition accuracy of {\it JWST}\cite{soummer14, lajoie16}. Simulations of the MIRI four quadrant phase mask contrast with sub-pixel grid dithers showed up to a factor of  ${\sim}10$ improvement over raw contrast \cite{soummer14}, and a factor of ${\sim}4$ improvement for NIRCam coronagraphic modes \cite{lajoie16}. Being able to test this observing strategy on-sky before the launch of {\it JWST} provides a useful test of this approach.

Sub-pixel dithering relies highly on the accurate centroiding of all the images. For STIS, this is straightforward. Because the masks and Lyot stop in the instrument are not optimized, significant diffraction spikes from the telescope support structure remain in coronagraphic images. While this is not ideal for imaging purposes, it does serve as a unique external probe of the stellar centroid. The diffraction spikes are nearly at $45^\circ$ angles from the vertical direction on the detector. Therefore, it is possible to measure the vertical location of the spike on the detector as a function of detector column and linearly fit the spikes to infer a stellar centroid. If one subtracts off the PSF halo, and uses a mask that encompasses only the diffraction spikes, accurate centroids can be obtained with this method, typically with uncertainties of better than $0.05$ pixel, or $2.5$~mas. Another centroiding method is the Radon Transform--based line integral\cite{pueyo15}. For STIS, one can apply an angular separation dependent correction map to the spikes, which can weigh the contribution from different pixels, then perform the line integral along $45^\circ$ and $135^\circ$ only\cite{ren17}, and obtain similar results as the previous method.

\subsection{High Contrast Image Post-processing for Point-Source Detection}
A large number of STIS high contrast observations have been obtained with classical high contrast imaging methods that rely on the stability and repeatability of the stellar PSF over periods of several hours. Other post-processing methods exist as well, and it is useful to consider the benefits of each type of technique in the context of STIS observing. In this subsection we investigate constructing high contrast images of HD~39383 using both classical PSF subtraction and Karheunen-Lo\`eve Image projection post-processing \cite{soummer12}.

\subsubsection{Classical MRDI Subtraction}
For all the 9 visits\cref{note:observation} of HD~38393 with STIS, in each visit, the HD~38393 images from the same dither position were aligned with the ``X marks the spot'' method\cite{schneider14} and registered to a common centroid, then median combined into a single image for the dither position. PSF subtraction for each dither position within each visit was done using adjacent orbits: for example, visit 4 used either the visit 5 or visit 6 image at the same dither point as a PSF reference. The better subtraction was determined by eye and a mask for each dither point was constructed. All nine separate spacecraft orientations were de-rotated and combined. This procedure is most reminiscent of MRDI, and is particularly well suited for STIS. In the case of Program 14426, our use of the host star as the reference guarantees that PSF subtraction residuals due to potential chromatic mis-matches in otherwise used template stars are fully eliminated. Otherwise, as discussed in Section~\ref{sec:hi-c}, if one uses a reference star, contrast performance may be degraded depending on how much the reference star departs from the target star's SED across the STIS bandpass.

\subsubsection{Karheunen-Lo\`eve Image Projection (KLIP) Subtraction}
The KLIP algorithm is used to reach the highest contrast to detect point sources\cite{soummer12}. To perform it onto the 810 total images (9 orientations $\times$ 9 dither positions $\times$ 10 CCD readouts), they were aligned by performing Radon Transform--based line integral using {\sc centerRadon} during a systematic post-processing for the entire 50CORON archive in Ref.~\citenum{ren18}. To minimize the influence from the BAR5 occulter and residuals from the diffraction spikes, masks were created on 87 $\times$ 87 pixel sub-arrays of the data and applied. For each sub-array image, we took the {\sc centerRadon}-determined positions of the star in the detector coordinates, then interpolated and cut a mask of the same size from the whole field-of-view STIS mask created in Ref.~\citenum{debes17}.

A reference library of PSFs were generated for each spacecraft orientation (90 images) from the other spacecraft orientations (720 images) to determine the contrast achieved within 24 pixel (${\sim}1\farcs2$): exterior to this angular separation, point-sources can be detected to the photon limits. Following the KLIP algorithm we generated model PSFs based on KLIP components of the STIS quasi-static PSF, and subtracted these from the target image (Section~\ref{sec:pointsource}). 

\subsection{Sensitivity to Point Sources}
\label{sec:pointsource}
To test the STIS contrast sensitivity performance of our HD~38393 observations with classical MRDI and KLIP, we inserted artificial point sources using calculated TinyTim PSFs\cite{krist11} \footnote{\url{http://www.stsci.edu/software/tinytim/}} using a G0 spectral type SED. To quantify the contrast limit for point source at different positions in the STIS images, and simulate real observations, only one point source was injected then reduced for SNR calculation. This process is then sequentially performed for point sources  at different physical locations on a grid in the polar coordinates: an angular separation, i.e., radius, ranging from $0\farcs2$ to $1\farcs2$ (4 pixel to 24 pixel) with a step of $0\farcs05$ (1 pixel), and an azimuthal angle from $0^\circ$ to $360^\circ$ with a step of $30^\circ$. We scaled the point sources relative to a source with the equivalent of a total aperture corrected $V$ band magnitude of 26.05 (${\sim}1$ e$^{-}$ s$^{-1}$ with STIS). To determine if a point source was recovered, we obtained aperture photometry with a $3{\times}3$ pixel square centered at its location, and determined whether the source was detected at an ${\rm SNR}\geq5$ through a binary search, with the background noise estimated from the other locations having the same angular separation as the injected point source to the central star. For each on-sky location, the iterative binary search start with two contrast limits, $10^{-3}$ and $10^{-8}$: in each step, we first inject a planet signal at that location with a trial contrast which equals to the average of current upper and lower limits, then we reduce the data and measure the SNR for that planet. If its SNR is greater than 5, then the new upper limit is set to that trial contrast; otherwise the new lower limit equals the trial contrast. We repeatedly update the upper or lower limits until when they converge, i.e., when the SNR for the trial contrast is within 0.01 from 5. To calculate the final contrast for a specific radius, we took the median of the contrasts obtained for all the azimuthal samples.

For KLIP, we also injected the interpolated TinyTim point source PSFs to simulate the dithered observations of one point source in all $9$ orientations as described at the beginning of this section. For the images from each orientation, the images from the other $8$ orientations were treated as references to construct the KLIP components; we selected the first $400$ components to reach the flat plateau of the KLIP residual variance\cite{soummer12} for a trade-off between maximizing quasi-static noise suppression and minimizing the over-fitting of random noise. For each injected point source, we used the $400$ components, we modeled the $90$ point-source--injected images in the other orientations and subtracted the model from the images. We then median-combined all the $810$ residual images and calculated the SNR for the point source. For a planet, we calculate its SNR assuming the small sample statistics\cite{mawet14}: we first obtain the signal by measuring the total counts of a 9 pixel box ($3{\times}3$) centered at the planet location and subtract it by 9 times the median of the pixels that are within 2 to 4 pixel from the planet. We then obtain the standard deviation for a ring of pixels that have stellocentric separation that are within ${\pm}1$ pixel of that of the planet (excluding the $3{\times}3$ region for the planet, consequently, a total of $n_{\rm ring}$ pixel), and multiply that number by $\sqrt{9+\frac{9}{n_{\rm ring}}}$ to obtain the noise for the planet. The SNR is thus the signal divided by the noise.

To calculate the contrast sensitivity curve with KLIP, we adopt the above injection and reduction procedure for one point source, and calculated the SNRs for the sources injected on the above-mentioned polar grid. We performed a binary search until an SNR of 5 is reached. For each radius, the contrasts at different azimuthal angles were median combined as in the MRDI scenario. 

\begin{figure}
\begin{center}
\includegraphics[width=0.6\textwidth]{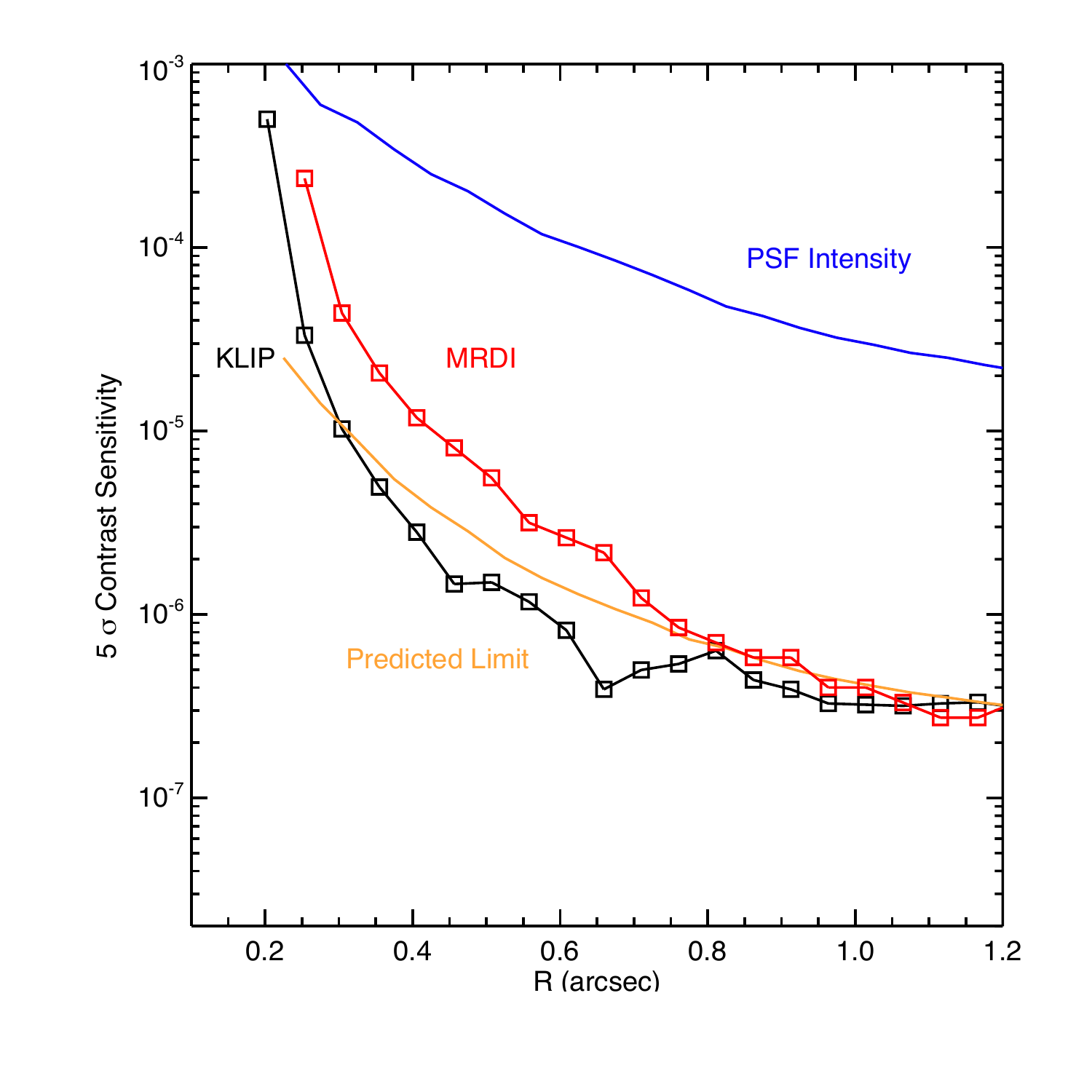}  
\end{center}
\caption 
[Contrast sensitivity curve for the special calibration Program 14426.]{ \label{fig:f6} Contrast sensitivity curve for the special calibration Program 14426. Azimuthally medianed point source sensitivity limits have been calculated as a function of radius from HD~38393, for classical MRDI reduction (red squares) and KLIP post-processing (black squares). These curves are compared to the PSF wing intensity divided by the peak stellar flux (blue curve) and the expected noise floor from both speckle noise and photon noise from the PSF wings combined (black curve) as well as photon noise from just the PSF wings (orange curve). The KLIP reduction, with the exception of the smallest angles, corresponds to the expected limit. MRDI performs as well as KLIP beyond 1$\arcsec$.
} 
\end{figure}

The contrast sensitivity curves of KLIP and MRDI, compared to predicted noise limits are presented in Figure \ref{fig:f6}. The figure shows the resulting point source sensitivities for the MRDI and KLIP reductions compared to the raw PSF wing intensity and the predicted noise model limited $5\sigma$ contrast $\mathcal{C}(r)$: 

\begin{equation}
\label{eq:e6b}
\mathcal{C} (r)= \frac{5 \sigma_{\rm tot}(r)}{C_{\rm ap}S_\star t_{\rm int}}
\end{equation}

where $\sigma_{\rm tot}$ was calculated following Equations \eqref{eq:e1} to \eqref{eq:e4}, assuming STIS parameters with $G$=4, $t_{\rm int}=162$ s, $t_{\rm exp}=0.2$ s, and $N_{\rm pix}=9$ for a $3{\times}3$ photometric aperture. A square aperture with nine STIS pixels results in an aperture correction $C_{\rm ap}=0.5$. Since we used the central star as the reference for MRDI, we assumed $A_{\rm ref}=A_{\rm t}=2$ for $\sigma_{\rm tot}$ (See Equation \eqref{eq:e4}). We further accounted for decreased coverage of certain radii by estimating how many spacecraft orientations participated at a given angular radius. This has the effect of degrading expected contrast sensitivity rapidly near the inner working angle with STIS, typically by decreasing the effective exposure time for a given location. If extreme sensitivity is required near the inner working angle it is advisable to directly account for how many times a particular location is directly observed. Figure \ref{fig:f6} shows that the contrast sensitivity beyond 0.7$\arcsec$ matches that expected for noise from the PSF wings alone, and that KLIP is very close to the theoretical limit under our assumptions of speckle statistics with the exception of right near the inner working angle.

\subsection{The effect of sub-pixel dithering and post-processing on contrast performance}

HD~38393 was observed in Program 14426 with a $3{\times}3$ dithering pattern in one pixel during a visit, with a dithering step of ${\sim}1/4$ pixel (${\sim}12.5$ mas). To explore the improvement of point-source contrast with sub-pixel dithering, we generated contrast curves for four groups; for each telescope orientation, 

1) Non-dithering: we randomly chose $9$ readouts without replacement from the central dithering location; 

2) $3{\times}3$--dithering: we randomly chose $1$ readout from each of the $9$ dithering locations; 

3) Horizontal-dithering: we randomly chose $3$ readouts from each of the $3$ central horizontal dithering locations (i.e., parallel to BAR5); 

4) Vertical-dithering: we randomly chose $3$ readouts from each of the $3$ central vertical dithering locations (i.e., perpendicular to BAR5).

\begin{figure}
\begin{center}
\includegraphics[width=0.6\textwidth]{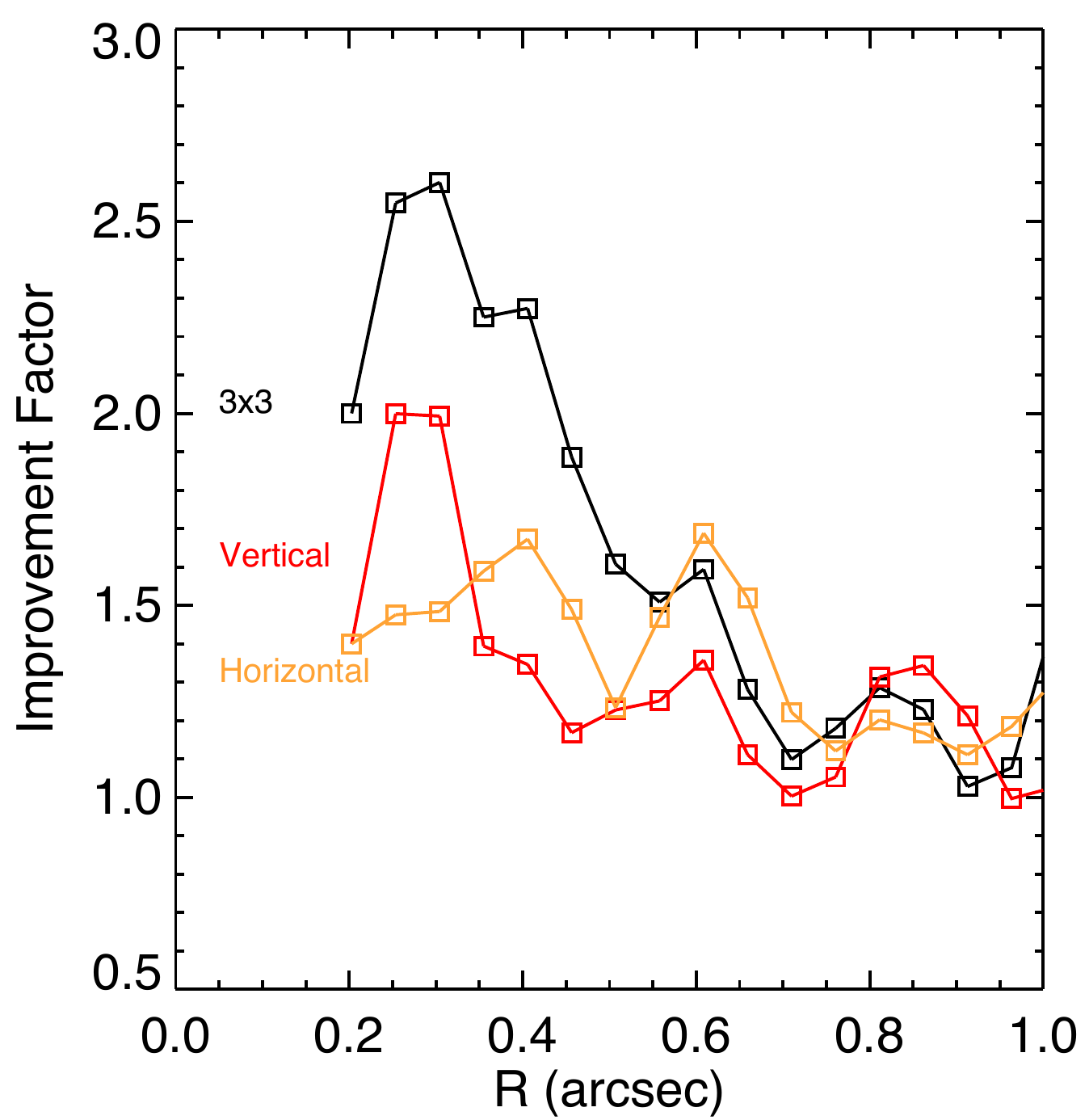}  
\end{center}
\caption 
[Contrast improvement through sub-pixel dithering with KLIP post-processing.]{ \label{fig:f7} Measured improvement of performance between different sub-pixel dithering methods and KLIP post-processing. We have investigated the performance improvement compared to no sub-pixel dithering for STIS. Dithering of any kind improves contrast, but a full $3{\times}3$ dithering approach appears to have the greatest impact, improving contrast by a factor of ${\sim}2$. 
} 
\end{figure}

For each of the four groups, there are a total of $81$ images ($9$ orientations $\times$ $9$ images), and the images from different dithering scenarios are matched to their dithering setups. We performed the identical KLIP point source contrast search as in Section~\ref{sec:pointsource}, but only using $40$ components to reflect the shrinking of the number of images. We first calculated the contrast curves for the four groups, then obtained the improvement factors over the non-dithering case by dividing the contrasts of non-dithered contrast by the dithered contrasts, and present them in Figure \ref{fig:f7}. 

From the improvement curves, we can see that both vertical dither and horizontal dithering improves the contrast limit, and a $3{\times}3$ (with a step size of ${\sim}1/4$ pixel, ${\sim}12.5$ mas)\footnote{This is the recommended pattern and step size in \url{http://www.stsci.edu/hst/stis/documents/isrs/2017_03.pdf}.} dithering strategy improves the performance by a factor of $\sim2$ at small inner working angles. These results are similar to those expected for the same strategy with {\it JWST}/NIRCam coronagraphic modes\cite{lajoie16}.

\subsection{Sensitivity to Circumstellar Material}
\label{sec:disk}

The observations of HD~38393 can also be used to gauge BAR5 performance in the detection of starlight scattered by circumstellar material. 

As in Section \ref{sec:pointsource}, we will determine STIS' sensitivity to spatially resolved disk-scattered starlight using artificial sources injected into the images and processed via KLIP. We assume a secure detection of disk-scattered starlight when the SNR in a given pixel is ${\geq}1$. This is acceptable when, by definition, a disk-signal is spread over at least several resolution elements (and typically much larger), so the SNR of the total disk-scattered starlight is ${\geq}5$.\footnote{For the extended structure in this paper, we have ignored the spatially correlated noise. Positively correlated noise will increase the noise level, our disk surface brightness limit is therefore optimistic. To reduce such noise, we used 1-pixel wide ring models. See Ref.~\citenum{wolff17} for a proper treatment of correlated noise using forward modeling.}  Thus we would require that any detectable disk-scattered starlight must have an ${\rm SNR}>1$ in ${\sim}25$ STIS pixels or an angular area of 6.4$\times10^{-3}$~arcsec$^2$. In practice, we calculate the signal to be the element-wise median of the derotated reduced images, and the noise to be the element-wise standard deviation of the derotated reduced images. 

We injected artificial disk-scattered starlight in the form of a circular face-on ring with a width of 1 pixel (thus unresolved in the radial direction, but avoids spatially correlated noise that impacts wider rings), and convolved it with a TinyTim point source STIS PSF, injected it to the 90 images of one orientation at HD~38393, then used the 720 images from the other 8 orientations as the references and performed KLIP subtraction; this identical process was performed for all of the 9 orientations, then derotated and median combined, which was used to calculate the average SNR in the ring. The actual observational design of the 14426 program is not well suited for the detection of disks at low inclination angles (i.e., nearly face-on), due to the lack of a PSF template and the use of MRDI. However, our approach efficiently determines STIS' azimuthally averaged sensitivity to spatially resolved disk-scattered starlight assuming a more advantageous experimental design is used, such as the use of a large reference library of PSFs from stars that do not possess extended disks.

\begin{figure}
\begin{center}
\includegraphics[width=0.6\textwidth]{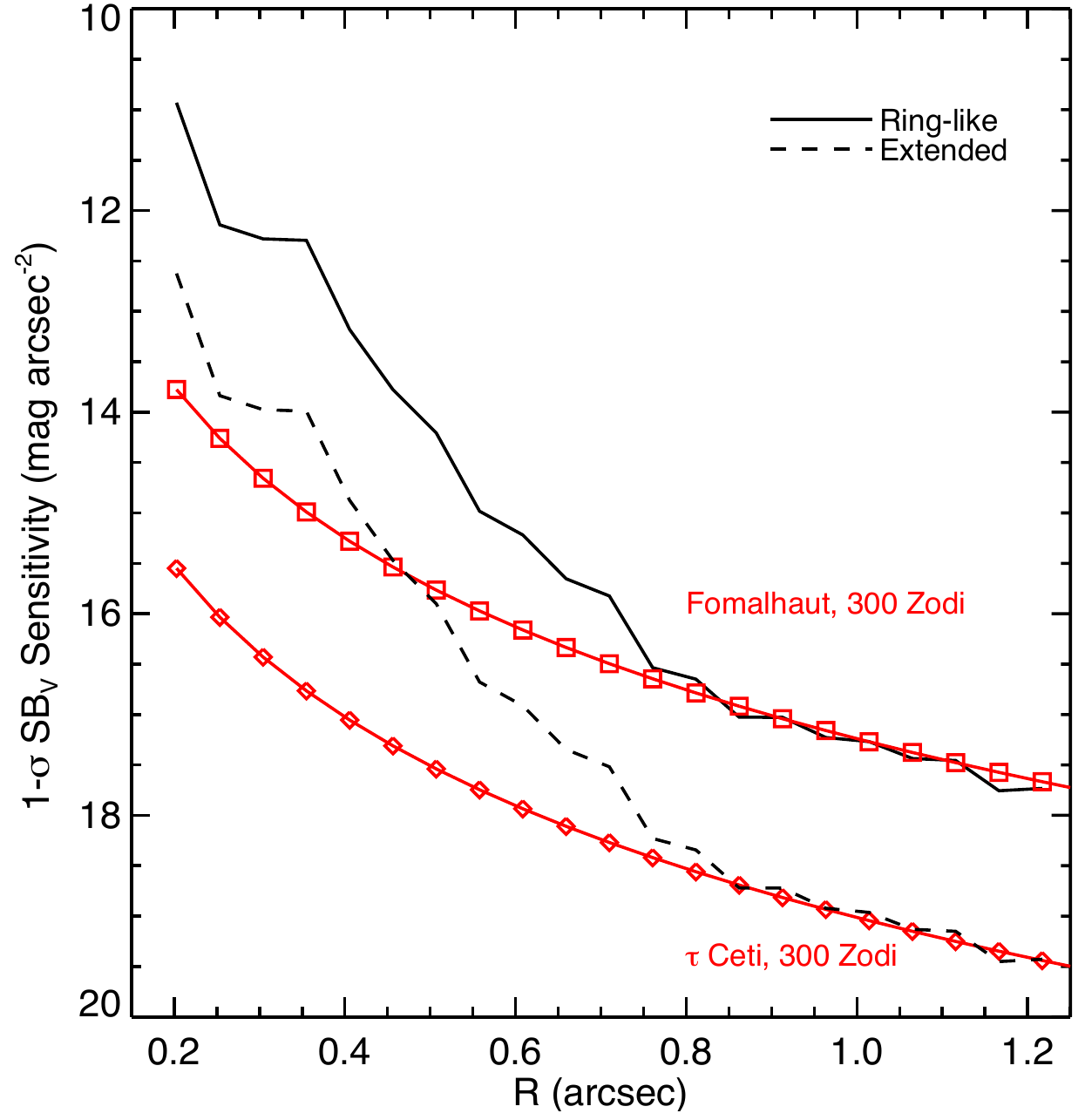}  
\end{center}
\caption 
[Expected sensitivity to circumstellar material for STIS+BAR5.]{ \label{fig:f8} Expected sensitivity to circumstellar material for STIS+BAR5. In this case we have implanted simple model disks and recovered them at an SNR of 1/pixel. The solid black line corresponds to a radially unresolved or narrow ring-like disk convolved with the STIS PSF, while the dashed black line corresponds to the limit of a fully radially resolved source that is not affected by convolution with the STIS PSF. Observed disks should reside between these limits. Overplotted we have calculated the surface brightness of hypothetical exozodi disks around Fomalhaut and $\tau$ Ceti, assumed to be fully resolved.
} 
\end{figure}

Assuming the scattering particles in the ring are spectrally neutral (i.e., grey scattering), we infer the scattered light sensitivities in units of $V$ band surface brightness in Figure \ref{fig:f8}. We show two possible sensitivity lines because the sensitivity is dependent on the convolution of the STIS PSF and the spatially resolved disk-scattered starlight. For radially unresolved disk-scattered starlight in a narrow ring-like disk (such as what we've modeled), the PSF washes the disk signal out by the enclosed energy fraction in the peak pixel. As the disk-scattered starlight approaches several resolution elements in total radial extent, the signal approaches the true surface brightness (the lower curve in Figure \ref{fig:f8}). This brackets the range of possible disk morphologies. We also have plotted the expected surface brightness of hypothetical exozodi around $\tau$~Ceti ($d=3.603\pm0.007$ pc\cite{gaia18}) and Fomalhaut ($d=7.70\pm0.03$ pc\cite{hipparcos07}). The exozodi disks are assumed to have constant optical depth, with their radii proportional to $L_\star/L_\odot$. The hypothetical models are $300$ times brighter than the Solar System zodical cloud at $1$~au (i.e., $V=22$ mag arcsec$^{-1})$\cite{stark}. As in Ref.~\citenum{stark}, we equate this surface brightness to 1 ``Zodi'', which is not to be confused with similar definitions that are based in the amount of infrared radiation from the Solar System's zodiacal cloud. We note that this result is for {\em predicted} performance with BAR5 and does not represent the actual sensitivity to scattered light for HD~38393 due to self-subtraction. That is beyond the scope of this paper.

\section{Recommendations for specific Scientific Cases}
\label{sec:cases}
Navigating the myriad aperture locations and formulating an optimal combination of them can be a fraught and confusing process. We thus try to detail some recommendations for common scientific cases with respect to high contrast imaging with the 50CORON aperture. Users are encouraged to consult with the HST helpdesk if their scientific case does not match those listed here. In general, if the scientific object of interest resides at 0\farcs25 $< r <$ 0\farcs5, it is recommended to use the BAR5 location. If the target resides beyond 0\farcs5, it is recommended to use the WEDGEA1.0 location as that has the most archival data available for post-processing libraries. For very low surface brightness objects beyond a $3\arcsec$ it is recommended to select the longest possible exposure times while still not saturating the detector at $\sim$2$\arcsec$ and to select the WEDGEB2.5 position\cite{ren17}, which has the most archival observations of the wider wedge positions. We point out particular HST GO programs that have utilized similar approaches to those detailed here--all Phase II programs are publicly available in MAST for users to study.

\subsection{Point Source at small inner working angles}

Let's assume that a young substellar object is the target of a STIS program and it has an SED similar to GL~229B (a T dwarf which is available as a non-stellar object option in the HST ETC). Its separation is 0\farcs3, and it is in orbit around an M5V star. From the STIS imaging ETC, such an object would have a contrast of 5$\times10^{-4}$ relative to its host in the 50CCD bandpass and thus would be detectable with STIS and the BAR5 position using an approach similar to Program 14426: ADI+post-processing. The user would need to use the noise model to assess the total exposure time needed, and would likely require at least three separate spacecraft orientations, taking care to select the ORIENT constraint in APT to place the companion perpendicular to BAR5's long axis. Assuming the PA of the companion is 30$^\circ$, ORIENTS of 75$^\circ$ and 255$^\circ$ would be optimal, based on the rule of thumb for the WEDGEB and BAR5 locations that one selects ORIENTS relative to a specific PA by adding 45$^\circ$ or 225$^\circ$. At this angular distance, one would select relative orientations of $>$20$^\circ$ to the nominal ORIENT to ensure that the companion was not significantly self-subtracted.

\subsection{Disk at moderate inner working angles with a known Position Angle}

Let's assume that the science target is an edge-on disk with a constant surface brightness of 18 mag arcsec$^{-2}$ beyond 1$\arcsec$ and its major axis has a PA of 125. In this case WEDGEA0.6 or WEDGEA1.0 would be appropriate depending on the brightness of the source and how quickly the PSF saturates at the inner working angle of each mask. In this case, one would select ORIENTS rotated 90$^\circ$ from the WEDGEB and BAR5 recommendation, since the WEDGA occulter is perpendicular to the other occulters' long axis. For users seeking to understand bright material close to the inner working angle yet also to detect faint nebulosity far from the target, a combination of WEDGEA0.6 and WEDGEA1.0 would provide larger dynamic range. This is similar to the strategy laid out in Program 12228. In addition to a selection of aperture location and orientation, it is paramount to obtain observations of a reference star with a similar magnitude, similar SED, and relatively nearby in the sky to ensure a similar thermal evolution of the observatory. If three orientations were planned, a fourth orbit would be needed to take observations of the reference star concurrent with the science target.

\subsection{Faint Point Source at large angular distance}
For this case we assume a very faint object with a contrast of 5$\times10^{-9}$ relative to a V=2 magnitude star at a separation of 9$\arcsec$, which is similar to Fomalhaut b, as studied by Program 12576 (PI: P. Kalas). In this case sensitivity and constrast is paramount over inner working angle and so the bright star is placed behind the WEDGEB2.5 aperture and allowed to saturate well beyond the full well. Even so, exposure times are short and of order 67~s per CR-SPLIT. Since the object is far from the star, small changes in spacecraft orientation for MRDI ensure that self-subtraction is not an issue for the point source target. MRDI and post-processing can be used to achieve the desired contrast level over twelve orbits.

\subsection{Face-on Disk}
In this case we assume observations of a near face-on disk that ranges in surface brightness from 14 mag arcsec$^{-2}$ at 0\farcs3 to 22 mag arcsec$^{-2}$ at 4$\arcsec$ where full angular coverage and photometric repeatability is desired. This is similar to GO 13753 (PI: Debes), which observed TW~Hya with a combination of BAR5 and WEDGEA1.0. In this case ORIENT constraints are not needed, but a careful scheduling of multiple orbits with ORIENT offsets are essential. In this case the combination of two aperture positions that are perpendicular to each other affords nearly full azimuthal coverage with just three spacecraft orientations close to the inner working angle (See Figure \ref{fig:f2}). In the case of face-on observations of fainter edge-on disks near the inner working angle, the reference star should be very close in SED and in apparent magnitude to ensure minimal speckle residuals.

\section{Implications for {\it WFIRST}/CGI Full Field of View Imaging }
\label{sec:wfirst}
One unknown feature of future space-based coronagraphs, such as the CGI on {\it WFIRST}, is the exact speckle behavior of these telescopes once they are launched. In this work, we have demonstrated that space-based observations with {\it HST}/STIS have residual speckle noise that is well characterized by a steeply varying function with stellocentric angle. To that end, we can use {\it HST} observations to investigate the potential performance of future space-based missions in the absence of active wavefront control. Such an investigation is important for two reasons. First, typical dark holes created by active wavefront control usually are limited by the size (actuator count) of their deformable mirrors, thus restricted to a range of angular separations that can be typically smaller than the full angular scale of an extended object, including most spatially-resolved protoplanetary and debris disks known today. Second, there exists a non-zero probability that the deformable mirrors used in such an instrument in space may become severely degraded or stop working without the opportunity for repair. Finally, investigating the speckle behavior of operating space-based observatories can better inform input models to wavefront control algorithms, particularly if a physical basis for the observed speckle evolution is determined. Understanding the potential performance at large angular distances, or in the absence of a dark hole is useful for defining new observing modes and science cases for future high contrast imaging, particularly for resolved circumstellar material at moderate angular separations from a star.

We investigate the possible performance of a CGI-like instrument without a dark hole. To do this, we rely on parameters from the noise and detector model of CGI developed to estimate its sensitivity to exoplanets\cite{nemati17}, discussed below and including our STIS-derived speckle variance. {\it WFIRST}/CGI will possess a similar primary mirror size, thermal properties that should be more stable than {\it HST}, and lower pointing jitter due to a low-order wavefront sensor and a fast steering tip/tilt mirror.  It is thus a reasonable and conservative approach to extrapolate  performance from STIS. We note that Ref.~\citenum{nemati17} assumed a systematic speckle noise source that depends on the total exposure time while we have demonstrated that uncorrected speckles combine proportional to the speckles within a single readout. This is primarily due to the fact that for STIS the time scale for speckles to change is fairly short, compared to the assumption that systematic speckle variations in CGI with wavefront correction will be as long or longer than the total exposure time on a source. This assumption will be straightforward to test during the CGI commissioning phase.

To do our calculation, we modify our existing noise model in Equation \eqref{eq:e1} to include the impact of the EMCCD excess noise factor, the gain of the EMCCD ($G_{\rm EM}$), and a clock induced charge noise term ($S_{\rm CIC}$) \cite{nemati17}:

\begin{equation}
\label{eq:e7}
\sigma^2_{\rm det}=N_{\rm pix}t_{\rm exp}\left[S_{\rm dark}+\frac{S_{\rm CIC}}{t_{\rm read}}+\left(\frac{S_{\rm read}}{G_{\rm EM}^2}\right)\right]. 
\end{equation}

Next, we re-calculate the expected count rate from the star on the detector, if unocculted, since we need to relate this to the estimated speckle noise and the predicted PSF wing intensity ($S_{\rm PSF}$). We will relate this to the flux of the peak pixel on the detector, just as we did in Equation \eqref{eq:e3}. 

\begin{equation}
\label{eq:e8}
S_{\rm PSF} = \pi \left(\frac{D}{2}\right)^2 F_{o}  \tau \eta I_{\rm peak} I(\theta),
\end{equation}
where $D$ is the primary mirror diameter, $F_{o}$ is the incident photons from a G0V spectral type star with an apparent V magnitude of 5, $I_{\rm peak}$ is the fractional enclosed energy of a point source within the peak pixel of the PSF, $\tau$ is the throughput of the telescope and instrument (optics, 575~nm filter with 10\% bandpass, obscuration of telescope pupil), $\eta$ is the quantum efficiency (QE) of the detector, and $I(\theta)$ is the angular dependence of the PSF wing intensity. Table \ref{tab:t3} lists relevant values for assumed CGI parameters used in our calculations.\footnote{Taken from \url{https://wfirst.ipac.caltech.edu/sims/Param_db.html}}

The exact design for the Hybrid Lyot Coronagraph (HLC)\cite{seo16} and its performance without a dark hole is not currently known, as the design has not been finalized and few models of the performance without deformable mirrors have been generated at large angular separations. In principle, the detector FOV diameter could be as large as $9\arcsec$, and CGI will likely take observations both in total intensity and polarized intensity light. We therefore assume that a) there will be an observing mode in total intensity light, the PSF intensity at angles beyond $1\arcsec$ will approximate those of STIS while conserving flux with a different plate scale ($p_{\rm CGI}=20$~mas; $p_{\rm CGI}/p_{\rm STIS}=0.156$), and the CGI PSF will be depressed by a factor due to the presence of a focal plane mask and Lyot stop (i.e., the hybrid lyot coronagraph or the shaped pupil coronagraph; $f_{\rm CGI}$). Thus, we assume $I(\theta)=0.156~I_{\rm STIS}(\theta) f_{\rm CGI}^{-1}$.

We estimate the suppression factor $f_{\rm CGI}$ based on HLC designs that approximate the range in far-field PSF wing intensity in the presence of no wavefront correction when compared with the azimuthally averaged STIS PSF wings (E. Cady, private communication). At $1\arcsec$, the PSF intensity is roughly a factor of $3$ smaller than STIS, while at larger radii the intensity is a factor of 10 fainter after accounting for differing plate scales. We thus investigate the STIS PSF suppressed by a factor of $3$ and $10$ to bracket the likely on-orbit performance. A significant uncertainty is to what degree the mirror roughness power spectrum density might have on the far field wings of the PSF\cite{brown91}.
Finally, we add our empirical prescription for speckle intensity, accounting for the fact that $I_{\star}$ in this case is equal to $S_{\rm PSF}/I(\theta)$.

\begin{figure}
\begin{center}
\includegraphics[width=0.6\textwidth]{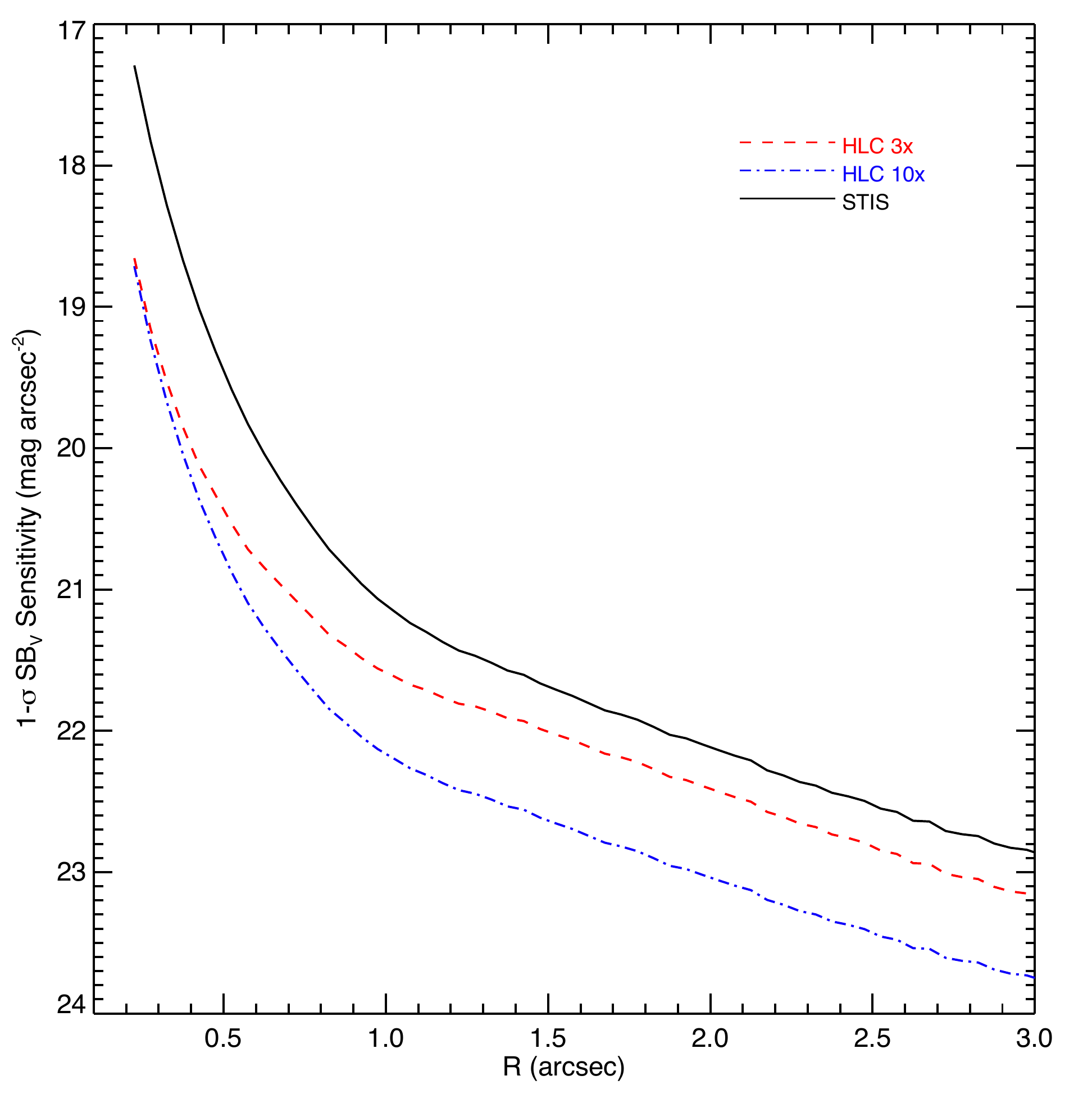}  
\end{center}
\caption 
[Performance comparison between WFIRST/CGI HLC and STIS.]{ \label{fig:f9} Comparison of the $1\sigma$ relative surface brightness sensitivity for 24~hr of observing time with 5~s readouts with STIS and hypothetical CGI/HLC performance assuming the PSF wings are 3x and 10x fainter than the STIS PSF wing intensity and high order wavefront correction is not used. Provided the HLC can suppress the PSF wings of the telescope to levels comparable or better than STIS, then far field sensitivity will be as good or better than current {\it HST} performance.
} 
\end{figure}

Figure \ref{fig:f9} shows the resulting $1\sigma$ surface brightness sensitivity in $V$ mag arcsecond$^{-2}$ as a function of radius between STIS' 50CCD bandpass and our hypothetical HLC performances for a $V=5$ solar type star. We find that CGI performance is expected to be better than STIS, despite the latter instrument's larger bandpass and platescale. The CGI performance at large angular radii can be further improved by binning pixels: the CGI spatial resolution is slightly oversampled by the detector in the nominal $575$~nm HLC band, and noise from binning should be negligible. Similarly, STIS observing efficiency with short exposure times is quite low due to spacecraft buffer management, Earth occultations, and detector readout, so {\em effective} additional performance can also be improved by obtaining efficiencies much higher than HST. At an L2 orbit and with faster detector readout times, WFIRST/CGI is likely to achieve higher efficiencies.

Our initial investigation here has two take-aways. First, {\it WFIRST} CGI can reach or exceed {\it HST}/STIS visible light high contrast imaging beyond the nominal HLC dark hole provided the HLC suppresses the PSF in excess of the STIS coronagraphic wedges. Secondly, this is particularly important for the study of previously resolved large disks, since only a handful of objects have measurements of disk photometry in multiple visible wavelengths, particularly interior to 1$\arcsec$. CGI thus would have the ability to characterize a large number of bright circumstellar disks between $0\farcs1$ and $3\arcsec$ that have previously only been observed with STIS and/or ACS.

\section{The Advantages of STIS versus ground-based high contrast imaging}
\label{sec:advantage}
STIS retains unique capabilities despite its age relative to newer ground-based coronagraphs. Firstly, the broad bandpass of the STIS CCD ensures that for solar-type and early-type stars, the effective wavelength of the observations is shorter than 5800~\AA, complementing near-IR observations from the ground. Especially for circumstellar disks, most ground-based observations are currently executed by observing in a differential polarization mode, or with heavy post-processing that rarely retains low spatial frequency information. STIS in combination with multiple spacecraft orientations and roll-differential imaging with a reference star PSF retains the full information of an image in total intensity light. STIS also has an unrivaled isoplanatic field of view compared to most ground-based coronagraphs, extending to distances of up to 25-50$\arcsec$ from a star depending on the aperture location.

\begin{figure}
\begin{center}
\includegraphics[width=0.6\textwidth]{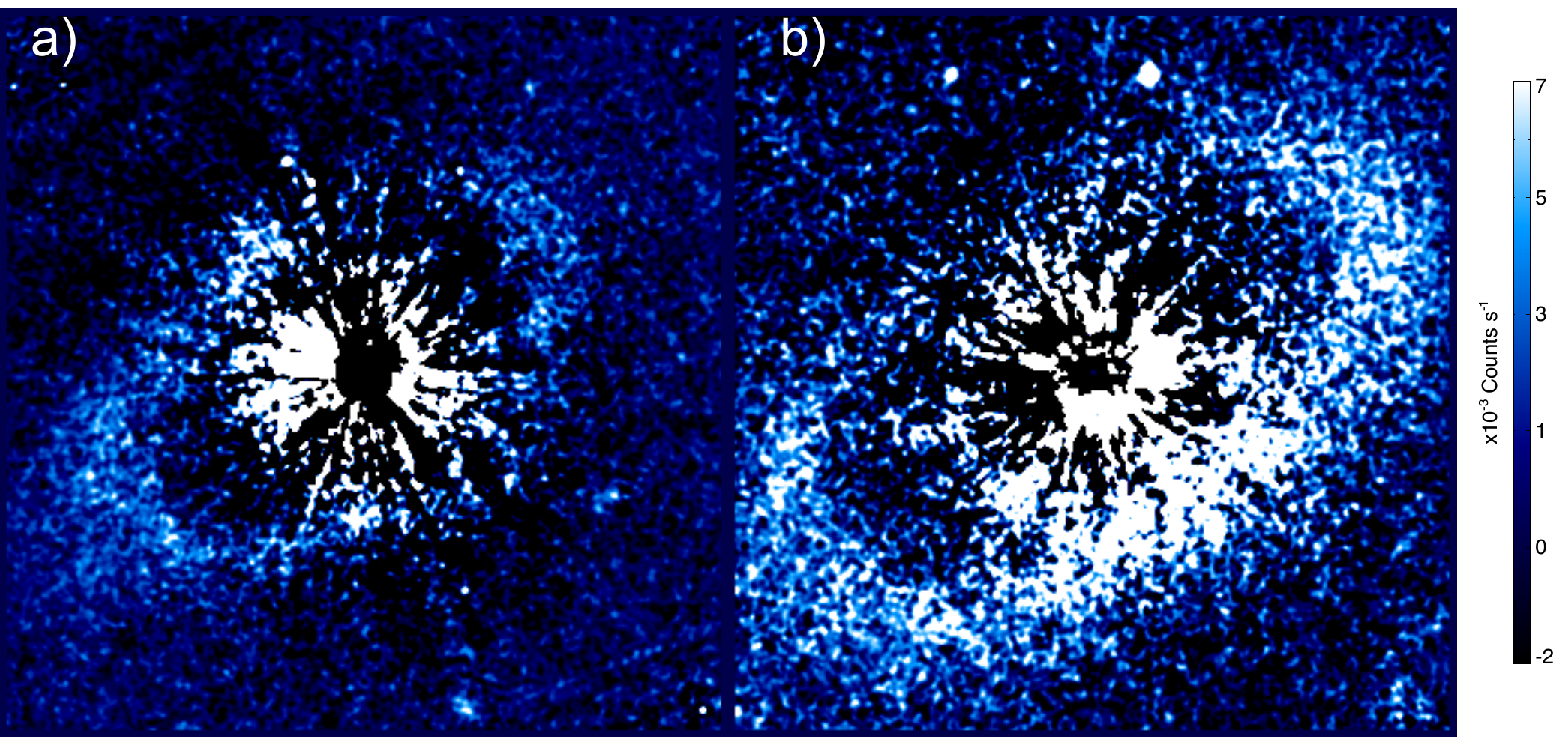}  
\end{center}
\caption 
[Low surface brightness debris disks.]{ \label{fig:f10} (a) Low Surface Brightness Debris Disk around HD~207129. HD~207129 is a solar type analog with a resolved debris disk first discovered by Ref.~\citenum{krist10}. Its peak surface brightness on the STIS CCD with gain $G = 4$ is ${\sim}4\times10^{-3}$ counts~s$^{-1}$ ($V\sim24.5$ mag~arcsec$^{-2}$)\cite{schneider17}. (b) Low Surface Brightness Debris Disk around HD~202628. HD~202628 is a solar-type analog with a resolved debris disk first discovered by \cite{krist12}. Its peak surface brightness on the STIS CCD with  gain $G = 4$ is ${\sim}2.5\times10^{-3}$ counts~s$^{-1}$ ($V\sim24$ mag~arcsec$^{-2}$)\cite{schneider17}. Images are smoothed by a Gaussian kernel with a 3 pixel FWHM.
} 
\end{figure}

STIS is well suited for very deep imaging in the visible for circumstellar material or faint companions, and compares favorably with ground-based observatories. An example is SPHERE/ZIMPOL images of TW~Hya's disk compared to those obtained with STIS, for similar total exposure times. Disk surface brightnesses fainter by a factor of a few hundred out to several arcseconds were detected in the STIS images compared to ZIMPOL\cite{debes17,vanboekel17}. Other examples include the detection of very low surface brightness debris disks such as HD~202628 and HD~207129 \cite{schneider16}. The $V$ band surface brightnesses of these disks are ${\sim}24.5$ and $24$ mag~arcsec$^{-2}$ respectively (See Figure \ref{fig:f10}). 

The stability of space also provides for very precise absolute flux measurements of high contrast images, which can be difficult to do from the ground due to constantly changing atmospheric conditions. An example can again be drawn from STIS images of TW Hya, which showed less than 10\% differences in azimuthally averaged disk surface brightness over a 15 year time period \cite{debes17}, or with $\beta$~Pictoris, which showed less than $2\%$ variation over 18 years\cite{apai15}.

Finally, STIS is not limited to stars brighter than a certain apparent magnitude, as is the case for AO-fed coronagraphs. There are practical limits to the faintness of a primary target, which depend somewhat on the desired contrast one wants to achieve and the exposure time for its acquisition. The limiting detectable magnitude of an object in fifty minutes of STIS integration time is roughly $V=27.3$\footnote{\url{http://etc.stsci.edu/etc/results/STIS.im.1162799/}}. For an object with $V=15$, the limiting contrast for a point source at most angular radii will be $10^{-5}$, due to the background sensitivity of the image. Observers should keep this in mind if they are observing faint primary targets. 

\begin{table}
\caption{\label{tab:t3} Assumed values for CGI performance.}
\begin{center}
\begin{tabular}{lcl}\hline\hline
Name & Value & Unit \\
\hline
Telescope Throughput & $0.5$  & \\
Detector QE ($575$~nm) & $0.929$ & \\
$I_{\rm peak}$ & $0.02$ & \\
Readnoise & $1.7\times10^{-6}$ & e$^-$  pixel$^{-1}$ frame$^{-1}$ \\
Dark Current & $7\times10^{-4}$& e$^-$  pixel$^{-1}$ s$^{-1}$\\
Clock Induced Charge & $3\times10^{-2}$ & e$^-$ pixel$^{-1}$ frame$^{-1}$\\
Platescale & $0.02$ & arcsec pixel$^{-1}$ \\
Primary Mirror Diameter & $2.37$ & m\\
Obscuration Fraction & $0.835$ & \\
\hline
\end{tabular}
\end{center}
\end{table}

\section{Conclusions}
\label{sec:conc}
High contrast imaging with STIS remains an active part of the unique observing modes of the Hubble Space Telescope. The recent commissioning of the BAR5 aperture location coupled with modern observing strategies and post-processing provides for high contrasts, small inner working angles, stable space-based PSFs, and deep sensitivities over a large FOV. This combination of capabilities over a total intensity visible bandpass allows complementary observations of objects with ground-based extreme AO-fed coronagraphs, {\it JWST}, and archival NICMOS data. 

\acknowledgments 
The authors would like to acknowledge the careful and painstaking work of A.~G\'asp\'ar in their commissioning of BAR5, and the leadership of the late B.~Woodgate in the design and construction of STIS without whom this work would not be possible. We also acknowledge useful conversations with B.~Nemati, B.~Macintosh, E.~Choquet, E. Douglas, L. Pueyo, and J.~Lomax as well as careful feedback from the two reviewers that aided in the clarity of this paper. GS acknowledges support from HST GO programs 13786 and 15219. This research has made use of the Direct Imaging Virtual Archive (DIVA), operated at CeSAM/LAM, Marseille, France. B.R.~acknowledges the computational resources from the Maryland Advanced Research Computing Center (MARCC), which is funded by a State of Maryland grant to Johns Hopkins University through the Institute for Data Intensive Engineering and Science (IDIES). 

\bibliography{report}   
\bibliographystyle{spiejour}   


\vspace{2ex}\noindent\textbf{John Debes} is an ESA/AURA Astronomer and STIS branch manager at the Space Telescope Science Institute.  He received a BA in Physics in 1999 at the Johns Hopkins University and a PhD in Astronomy and Astrophysics from the Pennsylvania State University in 2005.  He has authored or co-authored over 70 refereed papers and one book chapter.  His interests include high contrast imaging of exoplanets and debris disks, as well as the study of planetary systems that survive post-main sequence evolution.

\vspace{2ex}\noindent\textbf{Bin Ren} is a Ph.D.~candidate at the Department of Physics and Astronomy, and a M.S.Eng.~candidate at the Department of Applied Mathematics and Statistics, at the Johns Hopkins University. He received a BS in Physics and BEc in Mathematical Statistics at Xiamen University in 2013, and a MA in Physics and Astronomy at the Johns Hopkins University in 2016. His interests include post-processing methods for high contrast imaging of circumstellar systems, and characterization of them.

\vspace{2ex}\noindent\textbf{Glenn Schneider} is an Astronomer at Steward Observatory, and was the Project Instrument Scientist for HST’s Near Infra-red Camera and Multi-Object Spectrometer. His research interests are centered on the formation, evolution, and characterization of extrasolar planetary systems and have focused on the direct detection of sub-stellar and planetary mass companions to young and nearby stars and the circumstellar environments from which such systems arise and interact. He has led several STIS high contrast imaging programs, including the commissioning of the BAR5 aperture location.

\listoffigures
\listoftables

\end{spacing}
\end{document}